\documentclass[12pt]{article}

\usepackage{PRIMEarxiv}
\usepackage[utf8]{inputenc} 
\usepackage[T1]{fontenc}    
\usepackage{hyperref}       
\usepackage{url}            
\usepackage{booktabs}       
\usepackage{amsfonts}       
\usepackage{nicefrac}       
\usepackage{microtype}      
\usepackage{lipsum}
\usepackage{fancyhdr}       
\usepackage{graphicx}       
\graphicspath{{media/}}     

\usepackage{amsmath}            
\usepackage{titlesec}           
\usepackage{bm}                 
\usepackage{float}              
\usepackage{subfigure}          
\usepackage{caption}            

\title{The initial state of Pluto-Charon with Tidal Evolution
}

\author{
  Yun-Yan Lee \\
  Taiwan\\
  \texttt{d187108@gmail.com} \\
}

\begin{document}
\maketitle

\begin{abstract}
This study explores the gravitational interaction between Pluto and its moon Charon, which has led to their synchronous orbit, where they consistently show the same face to each other. This process is known as tidal evolution, which explains how the gravitational pull between two celestial bodies can adjust their spinning speeds until they match their orbital speed. By simulating various initial conditions, we aim to understand the specific paths Pluto and Charon might have taken to reach their current stage.  We observe that the tilt of their orbits has a negligible long-term effect, typically stabilizing quickly. In our analysis, we examine a scenario when the orbital size initially increases, leading to an increased orbit and a high eccentricity, a phenomenon attributed to the dissipation of tidal energy.
\end{abstract}

\section{Introduction}
In the Pluto-Charon system, tidal evolution significantly influences various orbital and rotational parameters, including the semi-major axis \(a\), orbital eccentricity \(e\), the spin angular velocities of Pluto \(\psi_p\), and Charon \(\psi_c\), as well as the orbital inclination \(I\). The end of tidal evolution is typically characterized by a spin-orbit resonance of $1:1$, where the spin velocity of the bodies matches their mean orbital motion, neglecting the effects of any permanent quadrupole moments. Currently, Pluto and Charon are observed to be in such a state, representing an end phase of tidal evolution.

In the context of the Pluto system, Charon's considerable mass relative to Pluto's other moons, such as Styx, Nix, and Hydra, positions it as a focal point for tidal evolution research. The minor gravitational influence of the smaller moons permits their exclusion from the detailed study without compromising the integrity of the analysis. The position of the barycenter outside of Pluto, in the space between Pluto and Charon, underscores the binary nature of their relationship. This aspect is critical for a comprehensive understanding of their joint tidal evolution, making the Pluto-Charon pair an exemplary subject for examining the principles of celestial dynamics within such systems.

Drawing on the work of \cite{canup2005giant,canup2010giant}, which employs smoothed particle hydrodynamic (SPH) simulations to investigate various potential initial states for the Pluto-Charon system, our study expands upon this foundation by examining a wide array of initial conditions.

The role of the permanent quadrupole moment is excluded from this study due to the established 1:1 spin-orbit resonance between Pluto and Charon. The permanent quadrupole moment can significantly alter the end state of tidal evolution, as demonstrated by Mercury's $3:2$ spin-orbit resonance with the Sun, contrary to the synchronous state that might be expected.

Our simulations aim to validate the findings against \cite{cheng2014complete}, enhancing the precision of the results, particularly in terms of adherence to the conservation of angular momentum. The simulations also extend to incorporate the effects of orbital inclination.

In the following context, Section 2 applies the $\Delta t$ model to investigate tidal evolution, inspired by observations of consistent final orbital semi-major axes for different initial eccentricities \cite{cheng2014complete}. This section utilizes equations from \cite{hut1981tidal} and \cite{zahn1977tidal} to simulate the evolutionary trajectory of the Pluto-Charon system and explore potential initial states. Section 3 examines the Q model to address the issue of orbital semi-major axis overshoot, to refine tidal evolution expressions based on insights from \cite{Zahn1966Les} and \cite{zahn1977tidal}. Section 4 introduces the aspect of orbital inclination, drawing on \cite{mignard1980evolution,mignard1981lunar}, and observes its rapid reduction, consistent with the current negligible inclination of Pluto and Charon, albeit with limited implications for determining initial states.

To facilitate the forthcoming discussions, Table 1 presents all necessary parameters for the study.
    
\begin{table}[!htbp]
    \centering
    \caption{variable table}
    \begin{tabular}{c c c}
    \hline
     variable & meaning & note\\ \hline
     G & gravitational constant &  \\ 
     $\dot\psi_i$ & spin velocity & i=p : Pluto; i=c : Charon \\ 
     $\alpha_i$ & spin velocity (mean motion) & \\
     $R_i$ & radius(km) & $R_p$:1153; $R_c$:606\\
     a &  orbital semi-major axis  &  \\
     X &  a/$R_p$  &  \\
     $M_i$ & mass(kg) & $M_p$:1.304e22 $M_c$:1.519e21\\
     $\mu$ & reduced mass & $\frac{M_p M_c}{M_p+M_c}$ \\
     $C_i$ & the moment of inertia ($Kg \cdot m^2$) & $C_p$:5.686e33; $C_c$:2.232e32\\
     $k_{2i} $ & Love number &  $k_2p$:0.058; $k_2c$:0.006\\
     $\triangle t_{i}$ & delay time  &  $\triangle t_p$:600\\
     $Q_{i}$ & dissipation function  &  $Q_p$:100\\
     I & inclination & $I_p=I$ $I_s=0$\\
     e & eccentricity & \\
     $A_{\triangle t}$ & the relative rate of tidal dissipation ($\triangle t$ model) & $\frac{k_{2c}}{k_{2p}}\frac{\triangle t_c}{\triangle t_p}(\frac{M_p}{M_c})^2 (\frac{R_c}{R_p})^5$ \\
     $A_{Q}$ & the relative rate of tidal dissipation (Q model) &
     $ \frac{k_{2c}}{k_{2p}}\frac{Q_p}{Q_c}(\frac{M_p}{M_c})^2 (\frac{R_c}{R_p})^5$ \\
    \end{tabular}
        
    \label{tab:my_label}
\end{table}

\newpage

\section{Tidal evolution with $\Delta t$ model}

Currently, the Pluto-Charon system has reached the end stage of the tidal evolution. The tidal interactions generate friction as each body induces tidal bulges on the other, with the bulges oriented along the line connecting their centers. This interaction results in a torque that either decelerates or accelerates its spin depending on whether a body is spinning faster or slower than its orbital motion, eventually leading to a state where Pluto and Charon face each other permanently, eliminating tidal friction.

Initially, we present the equations used for varying eccentricity orders, prefaced by a concise overview of tidal evolution. The concepts of $\Delta t$ and $Q$ arise from the lag angle between the spin velocity and orbital motion. For instance, when a satellite's spin velocity exceeds its orbital motion ($\dot\psi_p > n$), it generates a tidal bulge displaced by an angle $\delta$. In the $\Delta t$ model, $\Delta t$ correlates with $\delta=(\dot\psi_p-n) \Delta t$, whereas in the $Q$ model, $Q$ is related to $\sin 2\delta=-Q^{-1}$. These formulations imply that when the spin velocity synchronizes with the orbital motion, the lag angle vanishes, and consequently, tidal friction ceases.

In \cite[Chapter 5]{kopal2012dynamics}, a method is outlined for deriving the tidal evolution equation by identifying the perturbation forces and applying them within the framework of the Gaussian perturbation equations.

\subsection{The influence of different order at eccentricity}

Initially, we adopted the framework \cite{zahn1977tidal}, which offers a straightforward method to transition between the $\Delta t$ and $Q$ models. Nevertheless, this approach represents a simplified version, as it truncates the series expansion of eccentricity.
\begin{align}
\label{eqn: 123}
\begin{aligned}
     &\frac{1}{n}\frac{d}{d t}(\dot \psi_i)=\frac{3}{2}\frac{G M_j^2}{C_i R_i}(\frac{R_i}{a})^6  \\
    &\phantom{\frac{d}{d t}(I\Omega)} \cdot [\epsilon_2^{22}+e^2(\frac{1}{4}\epsilon_2^{12}-5\epsilon_2^{22}+\frac{49}{4}\epsilon_2^{32})+O(e^4)]\\
    &\frac{1}{a}\frac{d a}{d t}=-3n \frac{M_c}{M_p} (\frac{R_p}{a})^5  \\
    &\phantom{\frac{d a}{d t}}\cdot [\epsilon_2^{22}+e^2(\frac{3}{4}\epsilon_2^{10}+\frac{1}{8}-5\epsilon_2^{22}+\frac{147}{8}\epsilon_2^{32})+O(e^4)]\\
    &\frac{1}{e}\frac{d e}{d t}=-\frac{3}{4} n \frac{M_c}{M_p} (\frac{R_p}{a})^5  \\
    &\phantom{\frac{d e}{d t}}
    \cdot[\frac{3}{2}\epsilon_2^{10}-\frac{1}{4}\epsilon_2^{12}-\epsilon_2^{22}+\frac{49}{4}\epsilon_2^{32}+O(e^2)]
\end{aligned}
\end{align}

In this context, $\epsilon_2^{lm}$ is denoted in \cite{zahn1977tidal}, where we replace $\epsilon_2^{lm}$ with $k_2 \Delta t (l n - m \dot\psi_i)$, disregarding additional coefficients $E_n$ as outlined in \cite{zahn1977tidal}. Consequently, this substitution leads to evolution \eqref{eqn: 456}. Fig.1 illustrates the simulation outcomes derived from these equations.

\begin{align}
\label{eqn: 456}
\begin{aligned}
     &\frac{d\dot\psi_i}{d t}=-\frac{3G {M_j}^2}{C_i R_i}k_{2i}\triangle t_i  (\frac{R_i}{a})^6[(1+\frac{15}{2}e^2)\frac{\psi_i}{n}-(1+\frac{27}{2}e^2+O(e^4)]  \\
    &\frac{1}{a}\frac{d a}{d t}=\frac{6G}{\mu R_p^3}k_{2p}\triangle t_p {M_c}^2(\frac{R_p}{a} )^8[(1+\frac{27}{2}e^2)(\frac{\psi_p}{n}+A_{\triangle t}\frac{\psi_c}{n})   \\
    &\phantom{\frac{d X}{d t}=\frac{6G}{\mu X^7}k_{2p}\triangle t_p {M_s}^2(\frac{1}{R_p})^3+{}}-(1+23e^2)(1+A_{\triangle t})+O(e^4)] \\
    &\frac{1}{e}\frac{d e}{d t}=\frac{27G}{\mu R_p^3}K_{2p}\triangle t_p {M_s}^2 (\frac{R_p}{a})^8[\frac{11}{18}(\frac{\dot{\psi_p}}{n}+A_\triangle t \frac{\dot{\psi_s}}{n}) \\
    &\phantom{\frac{d e}{d t}=\frac{27e G}{\mu X^8}K_{2p}\triangle t_p {M_s}^2 (\frac{1}{R_p})^3}-(1+A_\triangle t)+O(e^2)]
\end{aligned}
\end{align}

\begin{figure}[!htbp]
    \centering
    \includegraphics[scale=0.2]{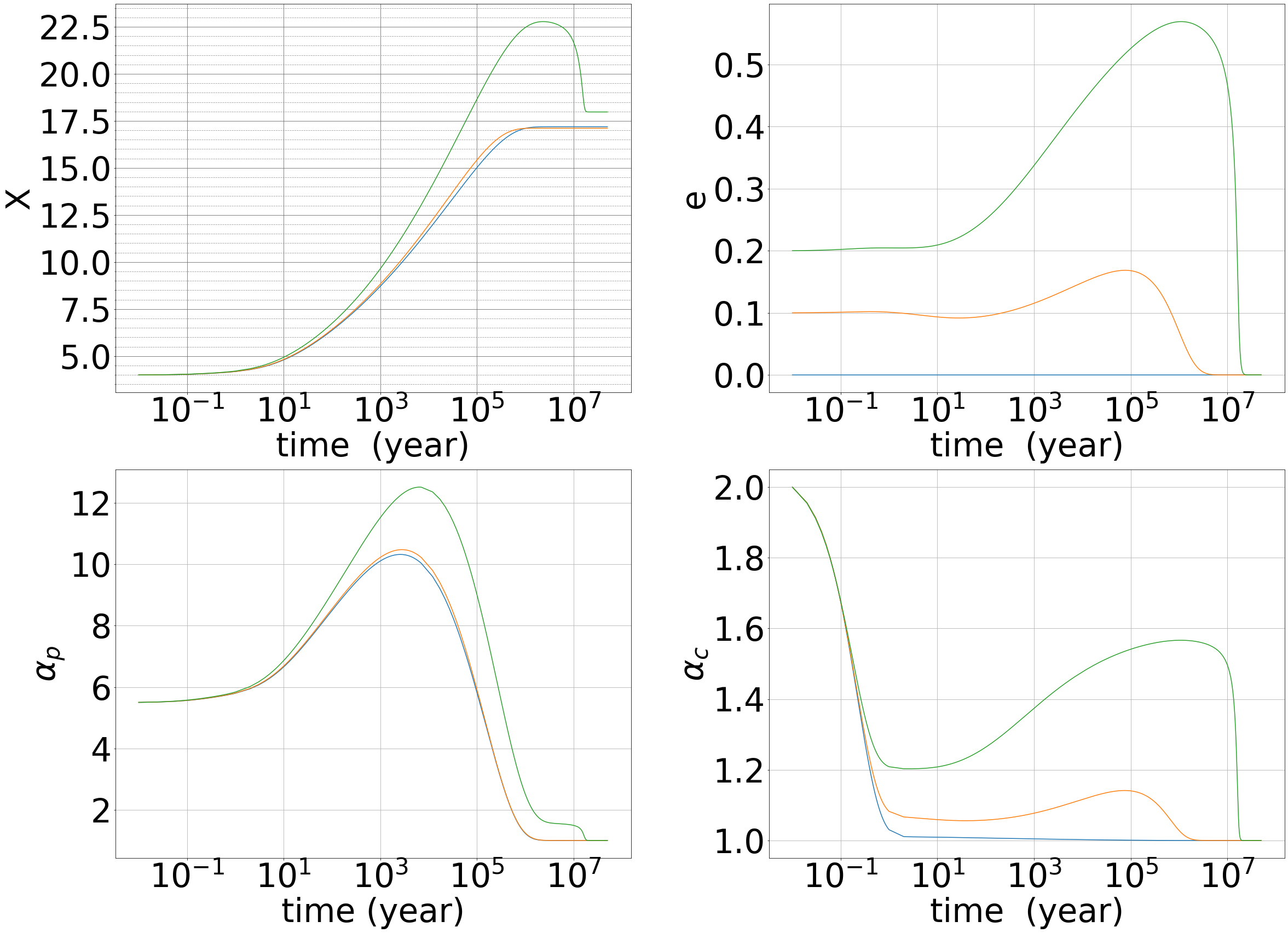}
    \caption{Tidal evolution with $\triangle t$ model follow the evolution \eqref{eqn: 456}}
    \label{Fig1}
\end{figure}

In Fig.\ref{Fig1}, the aim is to compare our findings with those presented in \cite{cheng2014complete}, utilizing identical parameters such as $A_{\Delta t}=10$ and analogous initial conditions. These include an initial semi-major axis to Pluto's radius ratio of $4$, Pluto's spin velocity to mean motion ratio $\alpha_p$ at $5.5$, Charon's spin velocity to mean motion ratio $\alpha_c$ at $2$, and initial eccentricities set at $0$, $0.1$, $0.2$, and $0.3$.

The outcomes depicted in Fig.\ref{Fig1} are derived from 
evolution \eqref{eqn: 456}, which accounts for eccentricity up to the second order. Notably, at an eccentricity of $e=0.2$, the orbital semi-major axis $a$ initially exceeds its eventual state before stabilizing, while at $e=0.3$, the system fails to reach a stable configuration. These observations are subsequently contrasted with the result in \cite{cheng2014complete}, which represents the $\Delta t$ model, to elucidate the discrepancies and challenges highlighted in Fig. \ref{Fig1}.


In \cite{cheng2014complete}, there is no observed overshoot in the orbital semi-major axis before it stabilizes at its current value, and his simulations show convergence even with an initial eccentricity of $e=0.3$. Intriguingly, adjusting the initial eccentricity does not affect the final state \cite{cheng2014complete}, a phenomenon he attributes to the lack of angular momentum conservation due to the truncation in the series expansion of eccentricity. However, this explanation is primarily associated with his $Q$ model analysis.

To address these complexities, we employed the comprehensive tidal evolution framework \cite{hut1981tidal}, represented by evolution \eqref{eqn: 789}. Fig. 3 showcases the outcomes of these simulations, which incorporate all orders of eccentricity. In \cite{hut1981tidal} the equation for tidal forces is truncated until the $r^{-3}$ term, neglecting the $r^{-4}$ perturbation term. This simplification is deemed acceptable as the impact of the omitted term is minimal compared to the significant effects of varying initial eccentricities.

\begin{align}
\label{eqn: 789}
\begin{aligned}
     &\frac{1}{n}\frac{d\dot\psi_i}{d t}=-\frac{3G {M_j}^2}{C_i R_i}k_{2i}\triangle t_i  (\frac{R_i}{a})^6[f_1(e^2)\frac{\dot\psi_i}{n}-f_2(e^2)] ,\\
    &\frac{1}{a}\frac{d a}{d t}=\frac{6G}{\mu R_p^3}k_{2p}\triangle t_p {M_c}^2(\frac{R_p}{a} )^8[f_2(e^2)(\frac{\dot\psi_p}{n}+A_{\triangle t}\frac{\dot\psi_c}{n}) ,  \\
    &\phantom{\frac{d X}{d t}=\frac{6G}{\mu X^7}k_{2p}\triangle t_p {M_s}^2(\frac{1}{R_p})^3+{}}-f_3(e^2)(1+A_{\triangle t})], \\
    &\frac{1}{e}\frac{d e}{d t}=\frac{27G}{\mu R_p^3}k_{2p}\triangle t_p {M_s}^2 (\frac{R_p}{a})^8[f_4(e^2)\frac{11}{18}(\frac{\dot{\psi_p}}{n}+A_\triangle t \frac{\dot{\psi_s}}{n}) ,\\
    &\phantom{\frac{d e}{d t}=\frac{27e G}{\mu X^8}k_{2p}\triangle t_p {M_s}^2 (\frac{1}{R_p})^3}-f_5(e^2)(1+A_\triangle t)],
\end{aligned}
\end{align}
where
\begin{align*}
\begin{aligned}
     &f_1(e^2)=(1+3e^2+\frac{3}{8}e^4)/(1-e^2)^{9/2}, \\
    &f_2(e^2)=(1+\frac{15}{2}e^2+\frac{45}{8}e^4+\frac{5}{16}e^6)/(1-e^2)^{6} ,\\
    &f_3(e^2)=(1+\frac{31}{2}e^2+\frac{255}{8}e^4+\frac{185}{16}e^6+\frac{25}{64}e^8)/(1-e^2)^{15/2}, \\
    &f_4(e^2)=(1+\frac{3}{2}e^2+\frac{1}{8}e^4)/(1-e^2)^{5}, \\
    &f_5(e^2)=(1+\frac{15}{4}e^2+\frac{15}{8}e^4+\frac{5}{64}e^6)/(1-e^2)^{13/2} .
\end{aligned}
\end{align*}

\begin{figure}[!htbp]
    \centering
    \includegraphics[scale=0.2]{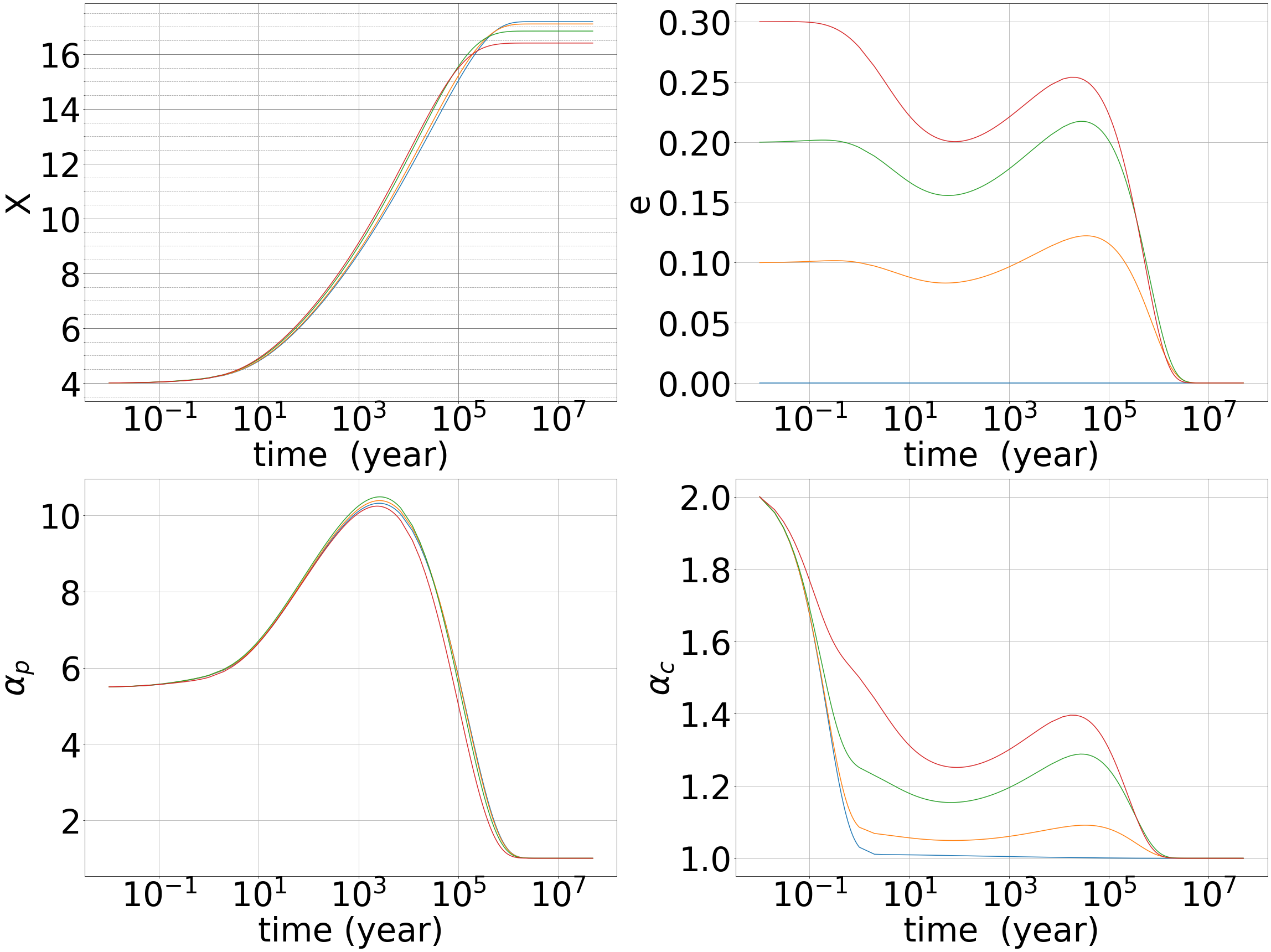}
    \caption{Tidal evolution with $\triangle t$ model follow the evolution \eqref{eqn: 789}}
    \label{Dt_Hut}
\end{figure}

Fig.\ref{Dt_Hut} demonstrates impeccable conservation of angular momentum and effectively addresses the issue of the orbital semi-major axis initially overshooting before stabilizing, as observed in Fig. 3. This consistency holds true even when varying the eccentricity, ensuring angular momentum is maintained throughout. Contrarily, adjusting the eccentricity does not impact the final orbital semi-major axis \cite{cheng2014complete}.

This subsection highlights the distinctions between the outcomes depicted in Figs. \ref{Fig1} and \ref{Dt_Hut}, providing valuable insights for refining the tidal evolution in the Q model. The implications of these findings and their application to model adjustments will be further explored in Section 3.

\subsection{The possible initial states evolve to the current state}
Our objective is to identify initial conditions that could lead to the current state of the Pluto-Charon system, with the conservation of angular momentum serving as a critical criterion for this determination. Therefore, we intend to examine various initial states to ascertain whether their tidal evolution culminates in the system's present configuration. Prior to evaluating these potential initial conditions, it is essential to ascertain the significance of different parameters in this evolutionary process, guided by the principle of angular momentum conservation.
\begin{align}
\mu\sqrt{G (M_p+M_c) a (1-e^2)}+C_p \frac{G (M_p+M_c)}{a^3}\alpha_p+C_c \frac{G (M_p+M_c)}{a^3}\alpha_c \label{10}
\end{align}
Given that $C_c \ll C_p$, the influence of $\alpha_c$ on tidal evolution is minimal. Additionally, the square of the eccentricity, $e^2$, remains significantly less than 1 for eccentricities of $0.1$, $0.2$, and $0.3$, indicating a minor impact on the evolution process. Consequently, our focus is to delineate a clear relationship that adheres to the conservation of angular momentum, particularly between the orbital semi-major axis and Pluto's spin velocity. To achieve this, we will introduce a wide array of initial conditions and identify those that converge to the current state of the system, as demonstrated in Fig.4. These specific points of convergence are verified through the application of evolution \eqref{eqn: 789}.


\begin{figure}[!htbp]
    \centering
    \includegraphics[width=\textwidth]{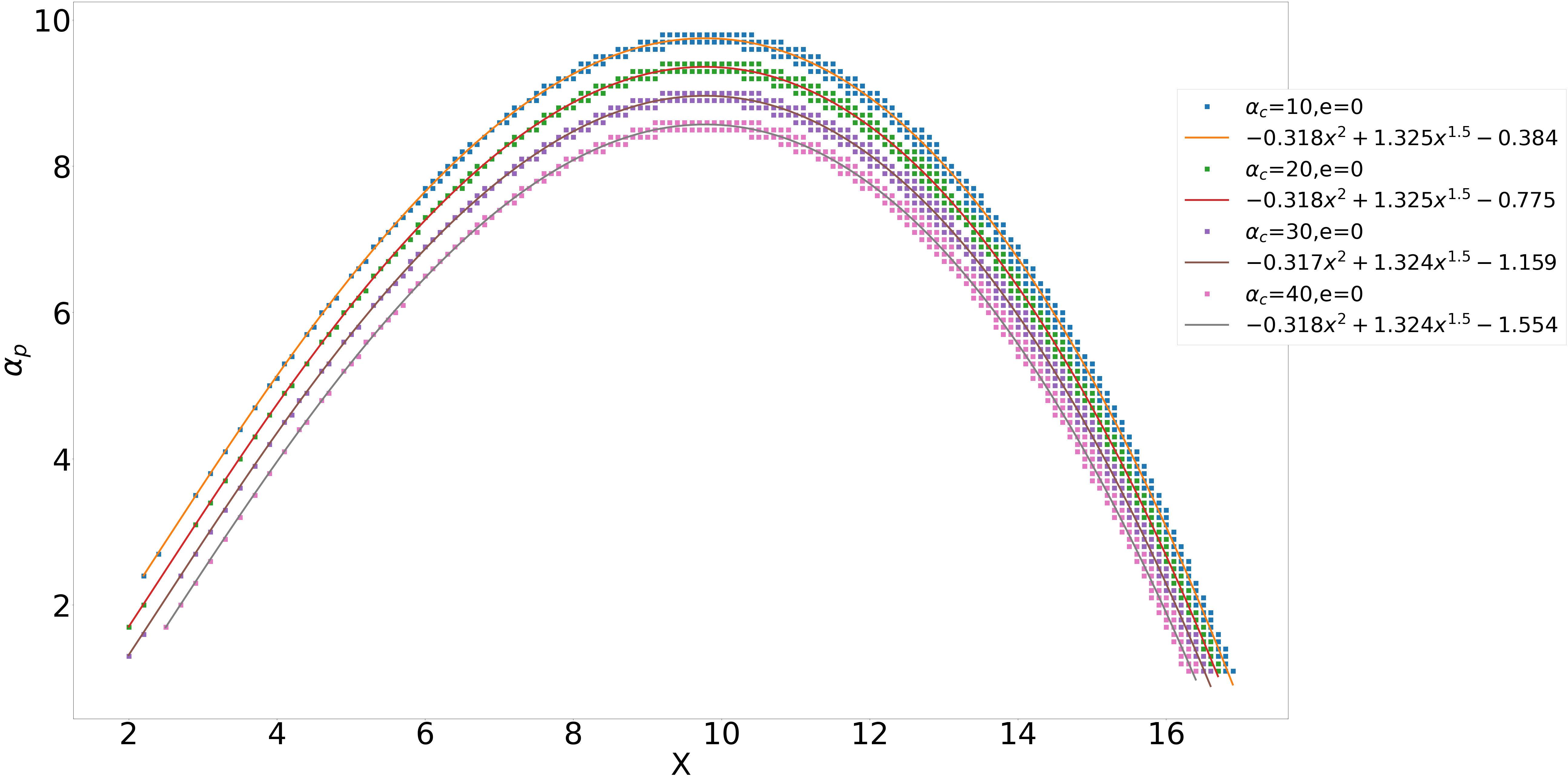}
    \caption{The possible initial state at $\alpha_c=10, 20, 30$, and $40$ and $e=0$}
    \label{fig: 4}
\end{figure}
In Fig. \ref{fig: 4}, the x-axis represents the orbital semi-major axis, while the y-axis corresponds to $\alpha_p$. For this illustration, the eccentricity is set to $0.0$, and various values of $\alpha_c$ ($10$, $20$, $30$, and $40$) are distinguished by different colors. The scatter points denote tested combinations of the orbital semi-major axis (ranging from $1$ to $17$ times Pluto's radius) and Pluto's spin velocity (from $1$ to $10$) that successfully evolve to the current state, defined as an end X within $16.9 < X < 17.1$. These points are plotted in Fig. \ref{fig: 4}, with the connecting line representing the fitted curve derived from the conservation of angular momentum equation, expressed as:
\begin{align}
\label{eqn: 11}
    \alpha_p = -\frac{\mu}{C_p} R_p^2 \sqrt{(1-e^2)} X^2 + \frac{\mu \sqrt{17} R_p^2 + C_c 17^{-1.5} + C_p 17^{-1.5}}{C_p} X^{1.5} - \frac{C_c}{C_p}\alpha_c
\end{align}

Setting $e=0$, we insert the values to obtain $\alpha_p = -0.318X^2 + 1.327X^{1.5} - 0.039\alpha_c$, which aligns with the fitted curves in Fig. \ref{fig: 4}.

\begin{figure}[!htbp]
    \centering
    \includegraphics[width=\textwidth]{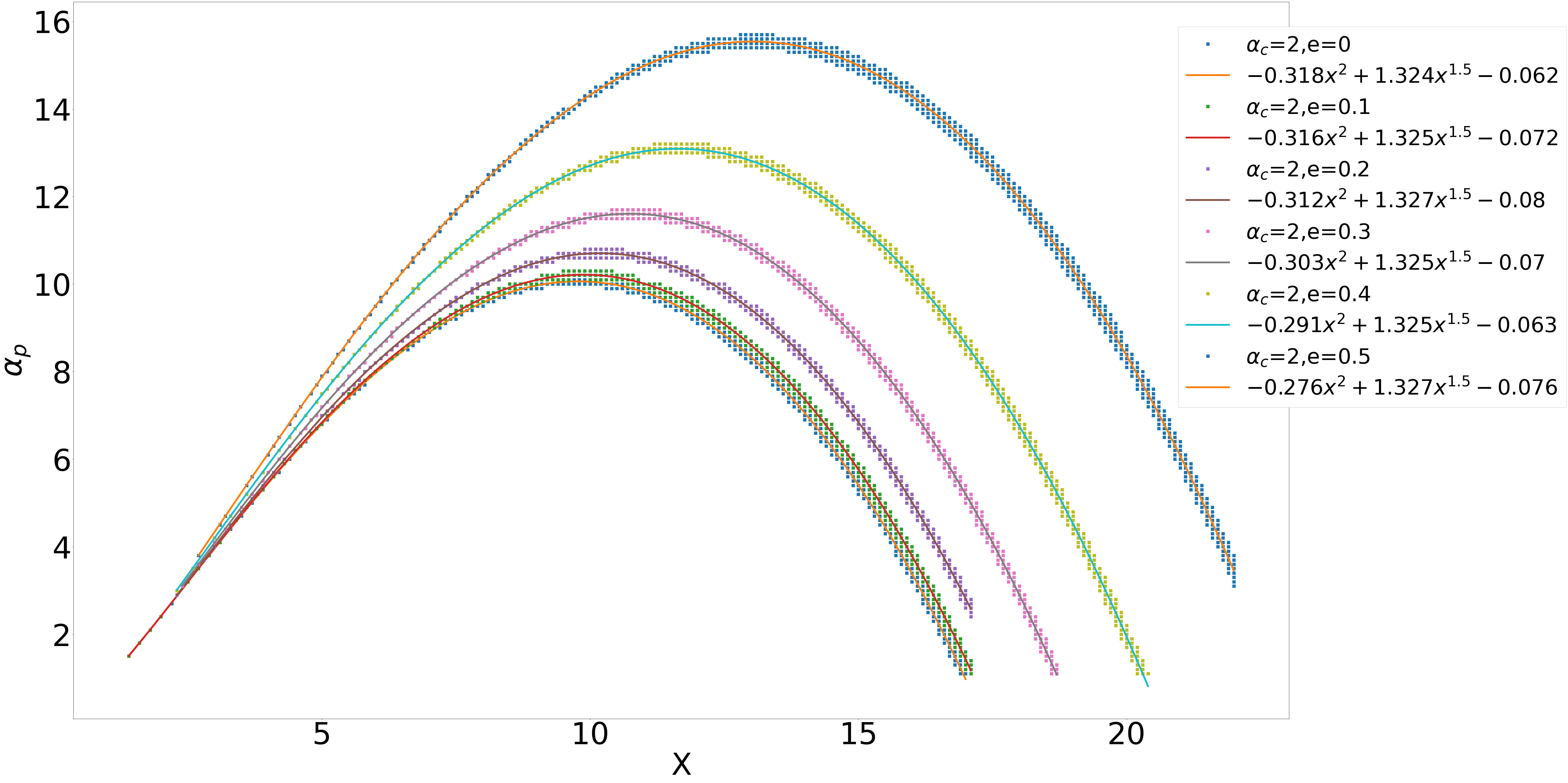}
    \caption{Illustration of potential initial states for varying eccentricities ($e=0, 0.1, 0.2, 0.3, 0.4, 0.5$) with Charon's spin velocity to mean motion ratio ($\alpha_c$) set at $2$.}
    \label{fig: 5}
\end{figure}

Fig.\ref{fig: 5} extends this analysis by illustrating the impact of varying eccentricities on the initial states.

In Fig.\ref{fig: 5}, an initial Charon's spin velocity to mean motion ratio ($\alpha_c$) of $2$ is established, with varying initial eccentricities explored, akin to the methodology in Fig.4. Here, the orbital semi-major axis range tested extends from $1$ to $22$ times Pluto's radius, and Pluto's spin velocity range from $1$ to $17$, to determine if these initial states can evolve to the current state, where the end X satisfies $16.9 < X < 17.1$. These qualifying points are then represented in Fig.\ref{fig: 5}, adhering to the relationship outlined in \eqref{eqn: 11}.

This approach underscores the pivotal role of angular momentum conservation in identifying feasible initial states for tidal evolution. By applying the principle of angular momentum conservation, we can systematically explore potential initial conditions. Further, plotting $\alpha_p$ against $X$ during the evolutionary process offers enhanced insight into the dynamics of tidal evolution, illustrating how these parameters interact over time to maintain angular momentum conservation.
\subsection{$\alpha_p-X$}
\begin{figure}[!htbp]
    \centering
    \begin{minipage}[t]{0.48\textwidth}
    \centering
    \includegraphics[width=8cm]{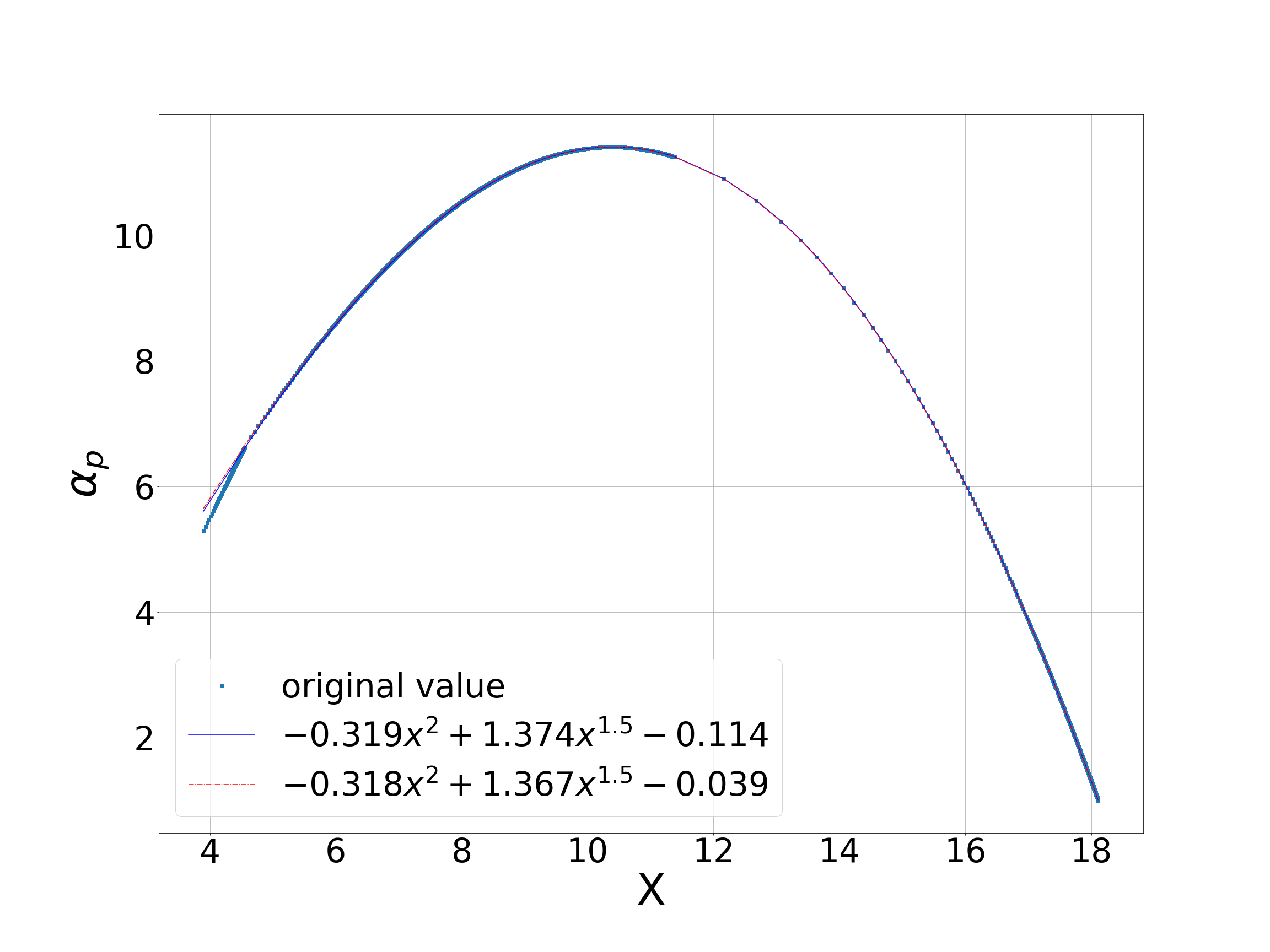}
    \captionsetup{justification=centering,margin=1cm}
    \caption{$\alpha_p$ vs $X$ at tidal evolution, and its initial state is $\alpha_p=5.3$, $\alpha_c=10$, $X=3.9$, and $e=0$.}
    \label{fig: 6}
    \end{minipage}
    \begin{minipage}[t]{0.48\textwidth}
    \centering
    \includegraphics[width=8cm]{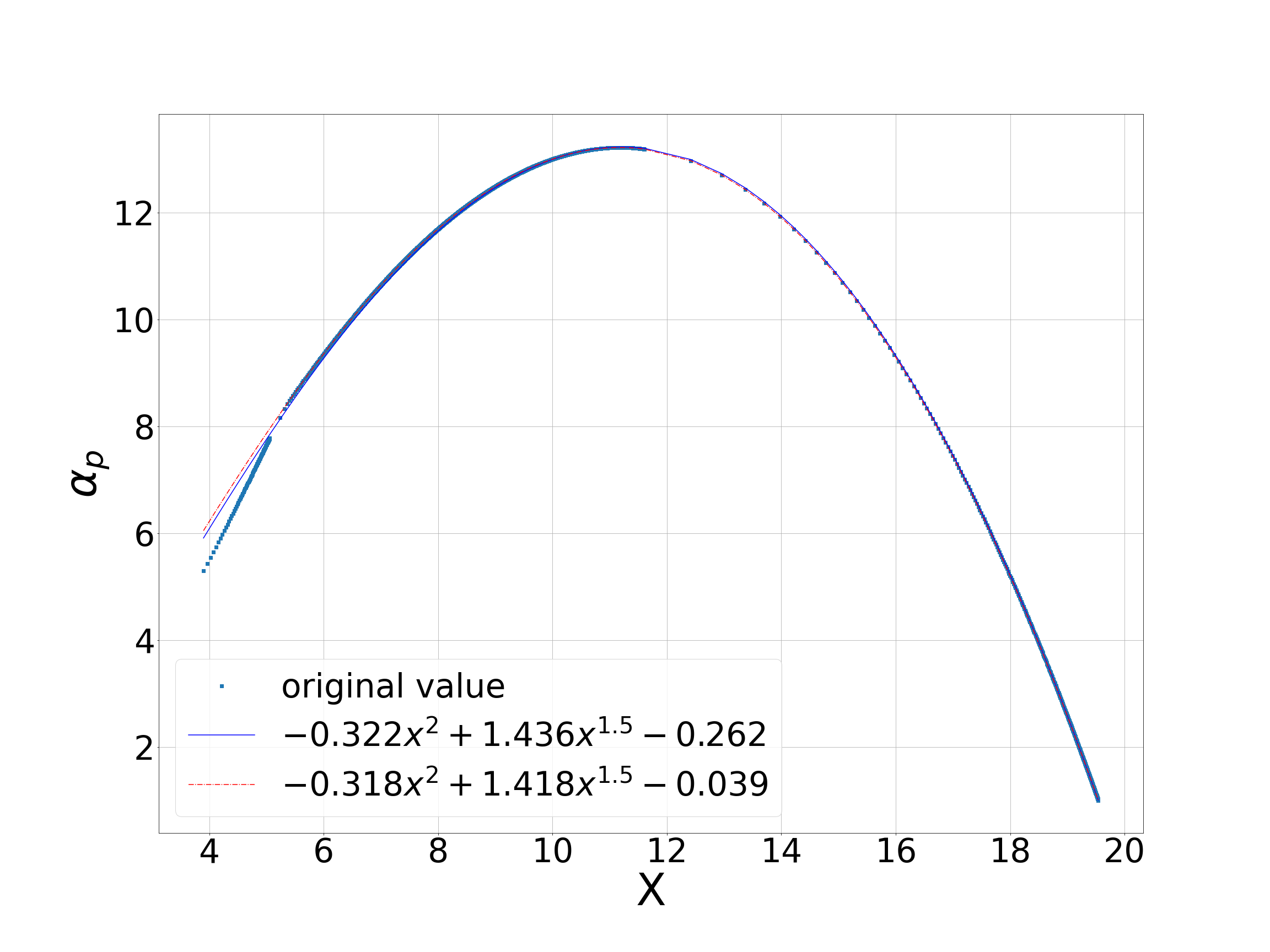}
    \captionsetup{justification=centering,margin=1cm}
    \caption{$\alpha_p$ vs $X$ at tidal evolution, and its initial state is $\alpha_p=5.3$, $\alpha_c=20$, $X=3.9$, and $e=0$.}
    \label{fig: 7}
    \end{minipage}
    \begin{minipage}[t]{0.48\textwidth}
    \centering
    \includegraphics[width=8cm]{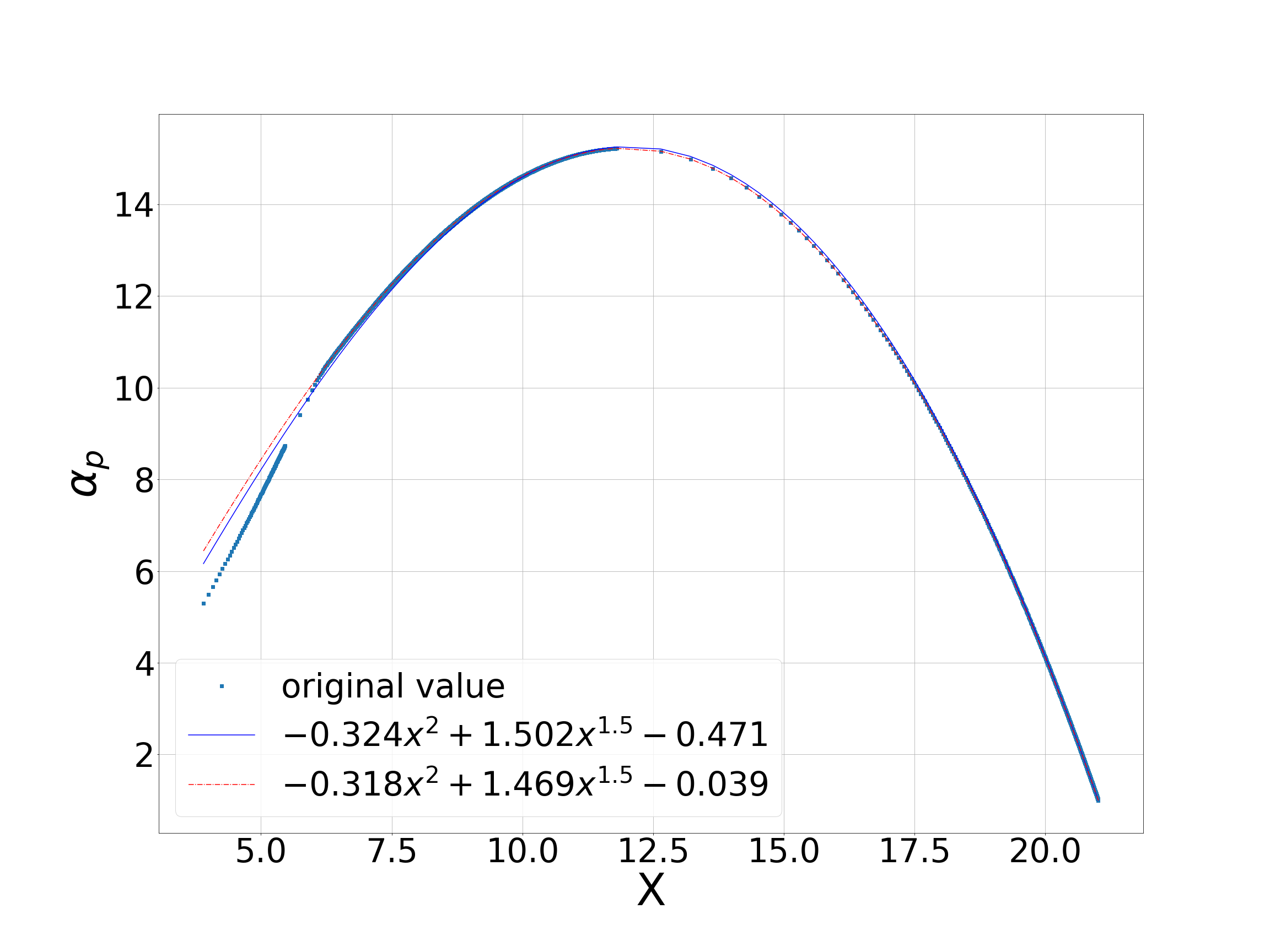}
    \captionsetup{justification=centering,margin=1cm}
    \caption{$\alpha_p$ vs $X$ at tidal evolution, and its initial state is $\alpha_p=5.3$, $\alpha_c=30$, $X=3.9$, and $e=0$.}
    \label{fig: 8}
    \end{minipage}
    \begin{minipage}[t]{0.48\textwidth}
    \centering
    \includegraphics[width=8cm]{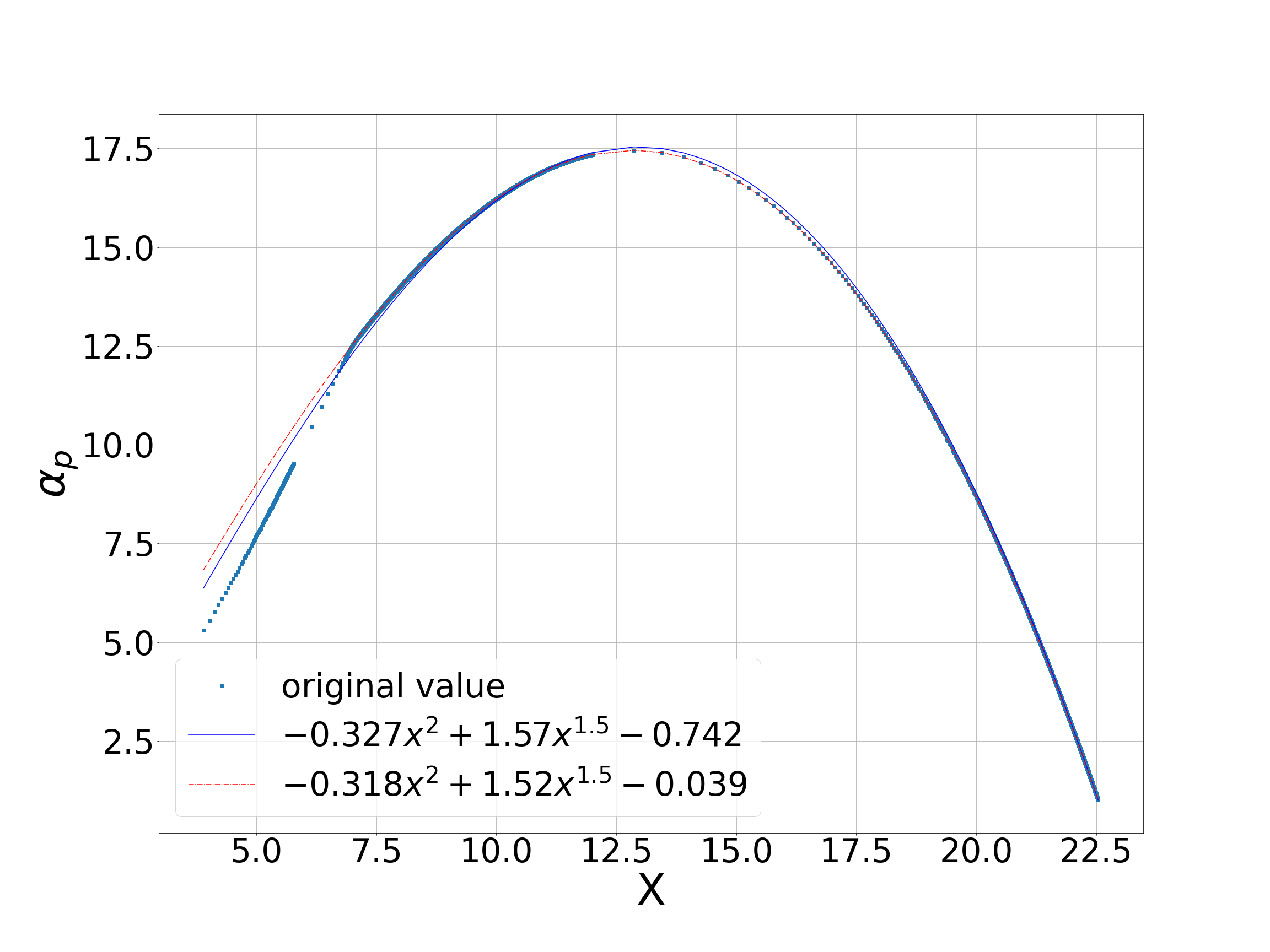}
    \captionsetup{justification=centering,margin=1cm}
    \caption{$\alpha_p$ vs $X$ at tidal evolution, and its initial state is $\alpha_p=5.3$, $\alpha_c=40$, $X=3.9$, and $e=0$.}
    \label{fig: 9}
    \end{minipage}
\end{figure}

In our simulations, an initial state characterized by $\alpha_p=5.3$, $\alpha_c=2$, a semi-major axis of $3.9$, and an eccentricity ($e$) of $0$ closely approximates the current state with $X=17$. Consequently, for illustrative purposes, we select an initial state with $\alpha_p=5.5$ and a semi-major axis of 4, as depicted in Figs. \ref{fig: 6} to Fig. \ref{fig: 17}.

Figs. \ref{fig: 6} to \ref{fig: 9} specifically explore variations in Charon's initial spin velocity.

Adjusting the initial Charon's spin velocity ratio ($\alpha_c$) leads to a rapid convergence towards $1$. In scenarios where the eccentricity is $0$, the Pluto-Charon system predominantly aligns with the red dashed curve, which represents the conservation of angular momentum based on the final state, particularly along the line where $e=0$ and $\alpha_c=1$. Here, the blue points denote results from the tidal evolution equations, with the blue curve as the fitted line.

At lower values of $A_{\Delta t}$, the deviation from the red dashed curve is more pronounced due to the significant impact of Charon's tidal forces on Pluto. The parameter $A_{\Delta t}$ dictates the magnitude of Charon's tidal influence on Pluto, hence at lower $A_{\Delta t}$ values, this effect is magnified, leading to a departure from the red curve. Conversely, higher $A_{\Delta t}$ values tend to prolong the system's alignment with the red dashed curve. The specifics of $A_{\Delta t}$'s impact will be elaborated on in the discussion section.

The system's prolonged adherence to the red dashed curve is analogous to the current Earth-Moon dynamics, where the Moon's spin velocity nearly matches its mean motion, and the eccentricity is minimal.

\begin{figure}[!htbp]
    \centering
    \begin{minipage}[t]{0.48\textwidth}
    \centering
    \includegraphics[width=8cm]{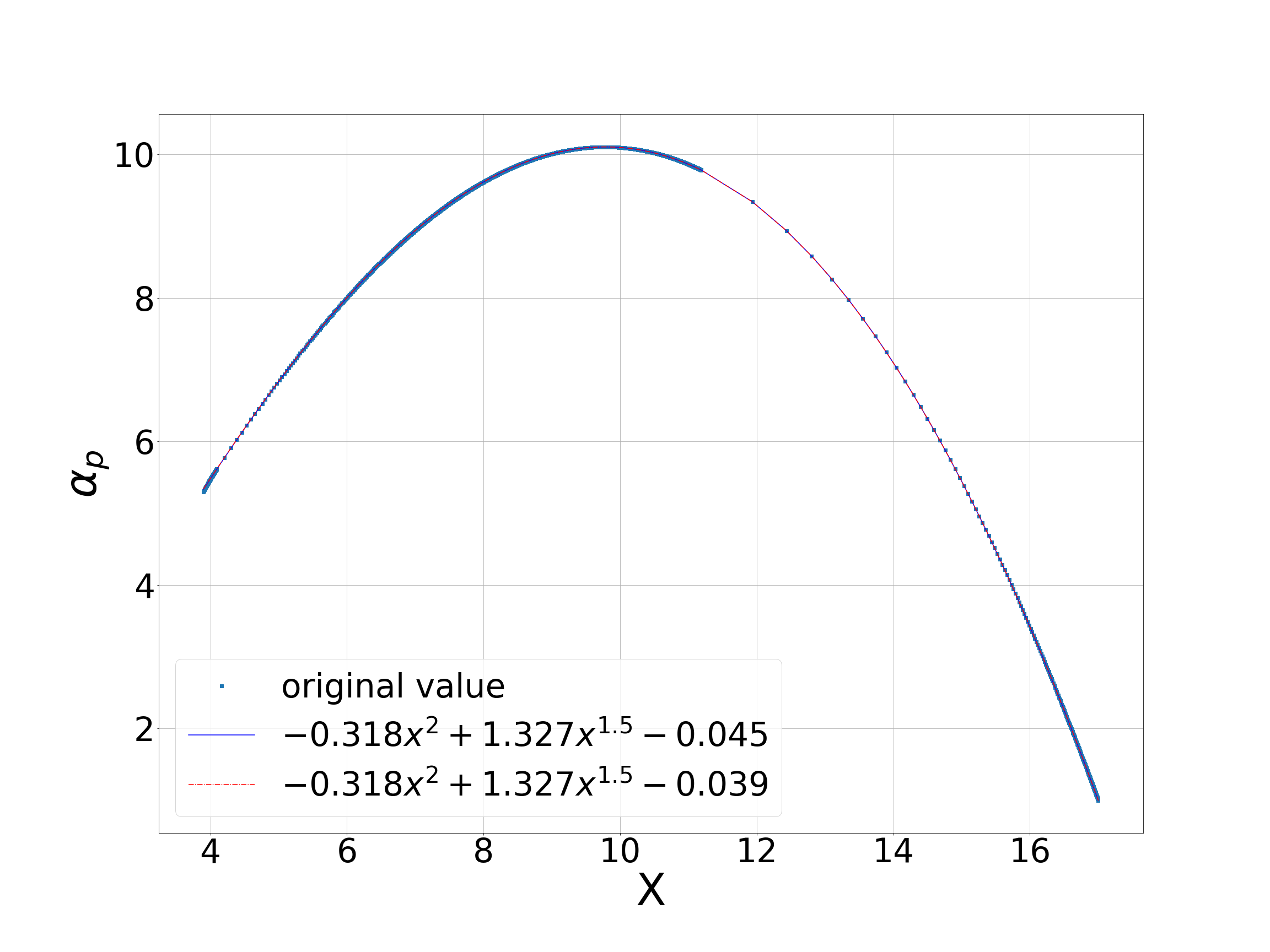}
    \captionsetup{justification=centering,margin=1cm}
    \caption{$\alpha_p$ vs $X$ at tidal evolution, and its initial state is $\alpha_p=5.3$, $\alpha_c=2$, $X=3.9$, and $e=0$.}
    \label{fig: 10}
    \end{minipage}
    \begin{minipage}[t]{0.48\textwidth}
    \centering
    \includegraphics[width=8cm]{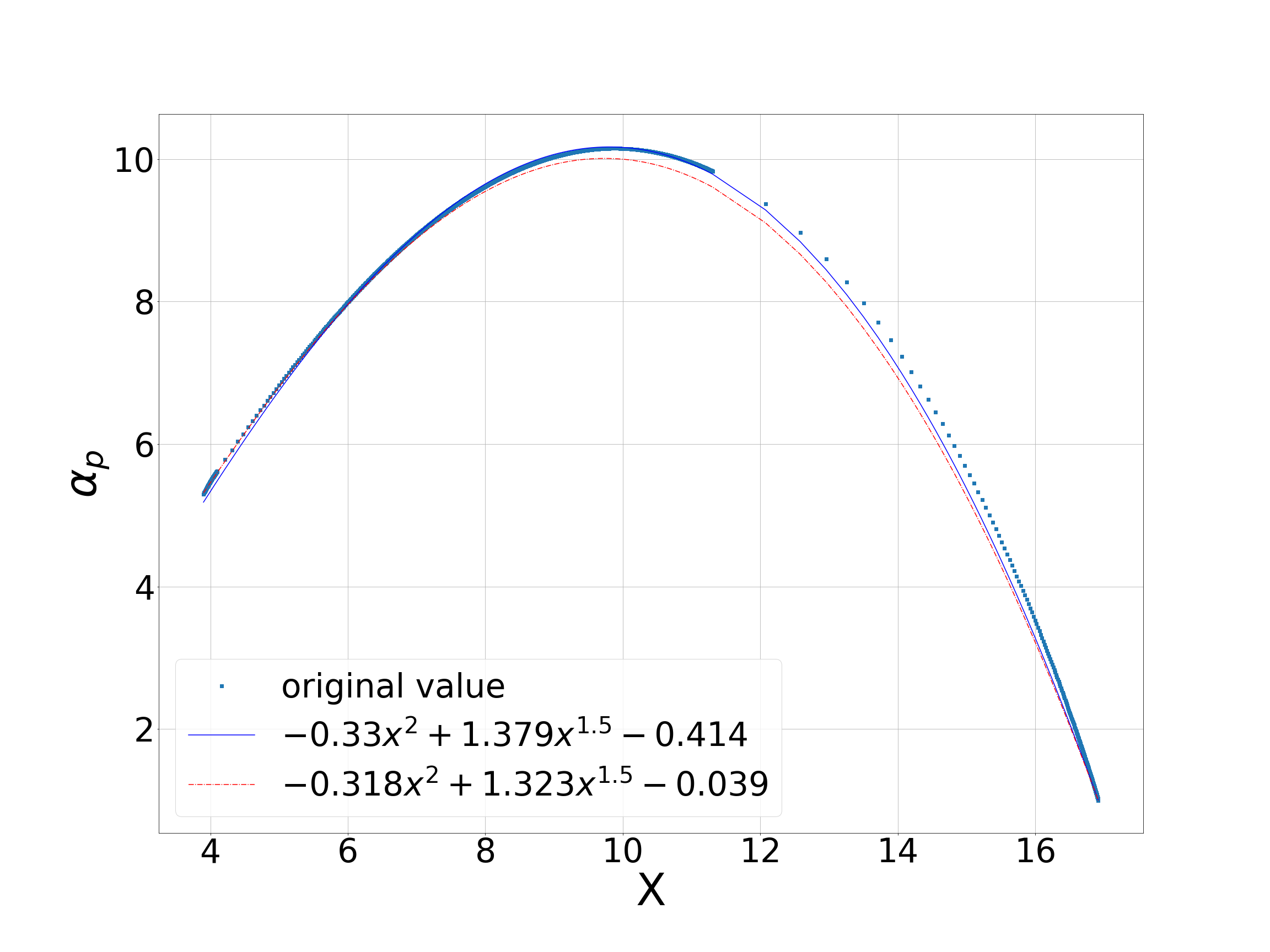}
    \captionsetup{justification=centering,margin=1cm}
    \caption{$\alpha_p$ vs $X$ at tidal evolution, and its initial state is $\alpha_p=5.3$, $\alpha_c=2$, $X=3.9$, and $e=0.1$.}
    \label{fig: 11}
    \end{minipage}
    \begin{minipage}[t]{0.48\textwidth}
    \centering
    \includegraphics[width=8cm]{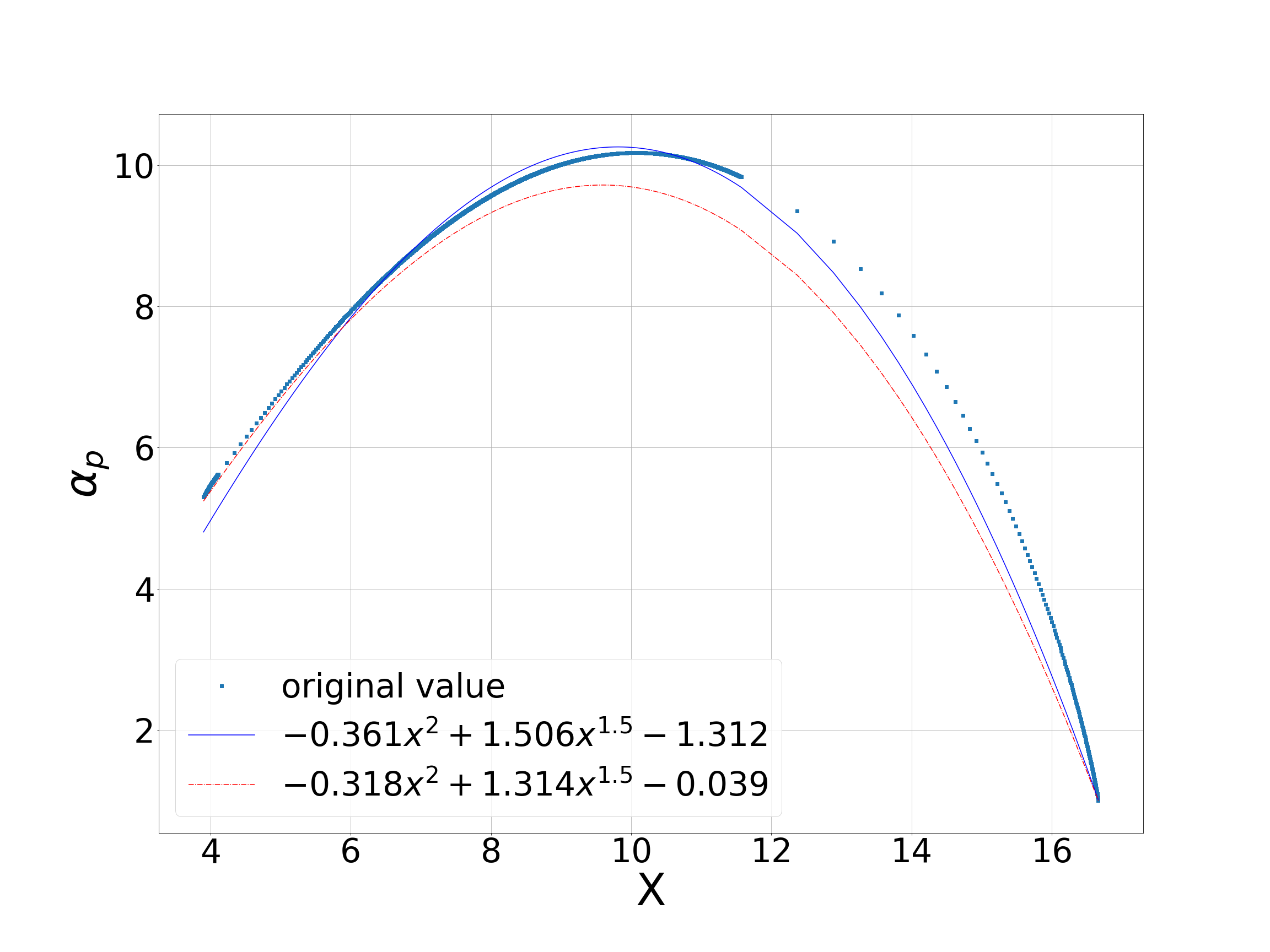}
    \captionsetup{justification=centering,margin=1cm}
    \caption{$\alpha_p$ vs $X$ at tidal evolution, and its initial state is $\alpha_p=5.3$, $\alpha_c=2$, $X=3.9$, and $e=0.2$.}
    \label{fig: 12}
    \end{minipage}
    \begin{minipage}[t]{0.48\textwidth}
    \centering
    \includegraphics[width=8cm]{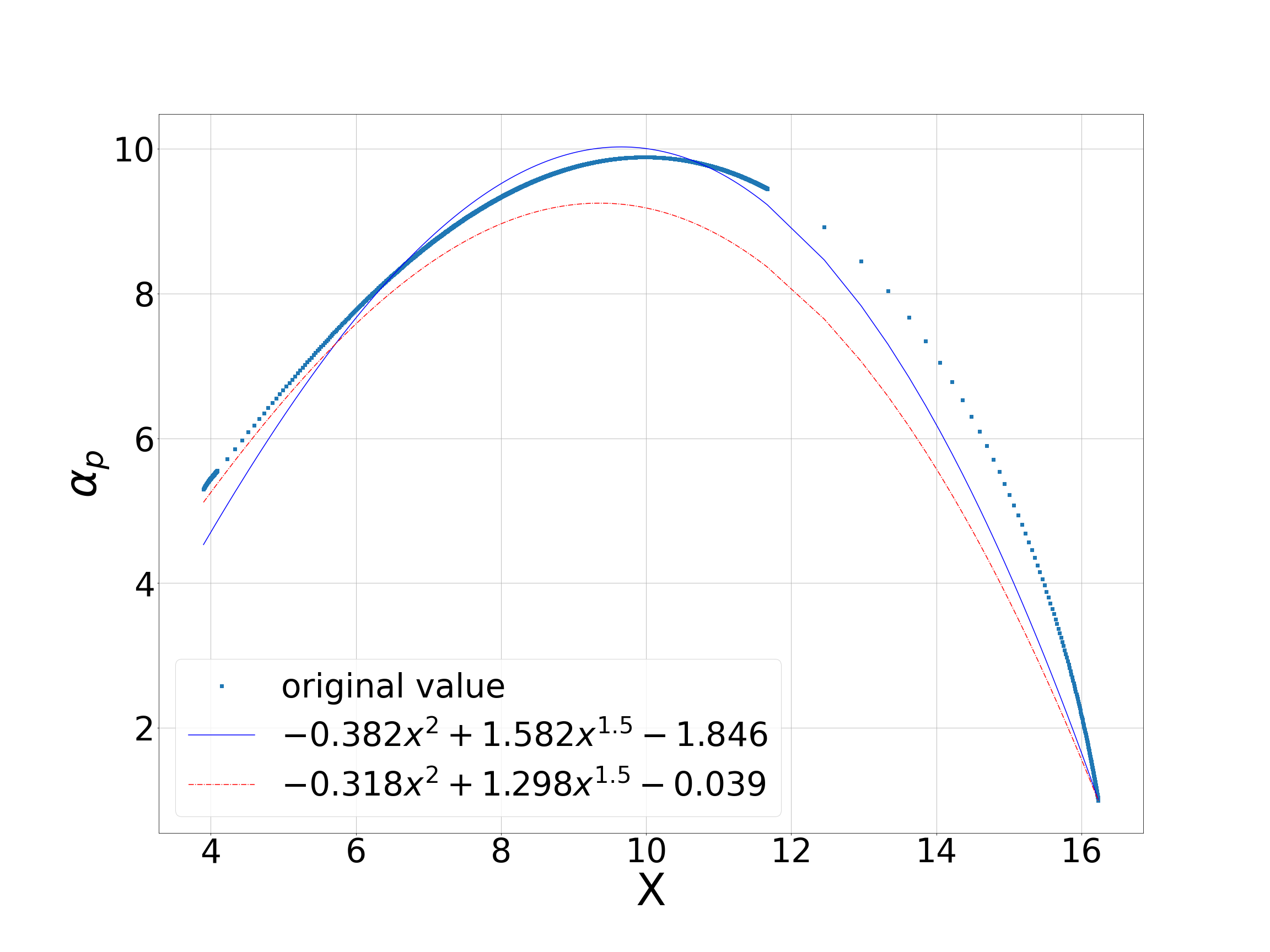}
    \captionsetup{justification=centering,margin=1cm}
    \caption{$\alpha_p$ vs $X$ at tidal evolution, and its initial state is $\alpha_p=5.3$, $\alpha_c=2$, $X=3.9$, and $e=0.3$.}
    \label{fig: 13}
    \end{minipage}
\end{figure}

Subsequent Figs. \eqref{fig: 10} to \eqref{fig: 13} examine the effects of varying initial eccentricities, presenting a contrast to the trends observed when adjusting Charon's initial spin velocity. Unlike in Figs. \eqref{fig: 6} to \eqref{fig: 9}, where a significant duration is spent at $\alpha_c=1$ and $e=0$, varying the initial eccentricity leads to a divergence from the red dashed curve indicative of $\alpha_c=1$ and $e=0$. This variation highlights the complex interplay between $A_{\Delta t}$ and the potential for overshoot, significantly influenced by Charon's tidal forces on Pluto, which will be discussed further.

Moreover, Figs. \eqref{fig: 10} to \eqref{fig: 13} reveal that higher initial eccentricities do not necessarily exert a greater influence on tidal evolution. These figures point to specific eccentricity ranges of particular interest, suggesting that while it may be challenging to pinpoint the exact initial state of Pluto-Charon through tidal evolution alone, this approach can help narrow down potential initial conditions for other celestial systems, especially if their current eccentricity is non-zero and their shape is sufficiently spherical to disregard the permanent quadrupole moment, thereby simplifying the search for their initial states due to the difficulty in achieving high eccentricity through tidal evolution.

\section{Q model-solving the overshoot}
Building on insights from Section 2, we observed that truncating the series expansion of the tidal evolution equation can lead to an "overshoot" of the orbital semi-major axis before it stabilizes at its current state. This issue is evident when comparing Fig. \ref{Fig1} with Fig. \ref{Dt_Hut}; Fig. \ref{Fig1}, based on a truncated equation \cite{zahn1977tidal}, exhibits this overshoot, especially in the Q model, whereas Fig. \ref{Dt_Hut}, adhering to a more comprehensive tidal evolution equation in the $\Delta t$ model, does not.

To address the overshoot issue observed in the Q model, we propose an approach akin to the $\Delta t$ model that incorporates all orders of eccentricity. Drawing from \cite{hut1981tidal,murray2000solar}, we define $\sin 2\delta = -\frac{1}{Q}$, focusing on the leading contributions to the tidal perturbing force to refine the Q model's accuracy and eliminate the unnecessary overshoot.
\begin{align}
    &F_r=-3 G \frac{m^2}{r^2} (\frac{R}{r})^5 k (1+3\frac{\dot r}{r} \frac{Q}{2\dot\psi-2\dot\theta}) \\
    &F_{\theta} = -3 G \frac{m^2}{r^2} (\frac{R}{r})^5 k \frac{Q}{2})
\end{align}
In this context, the energy perturbation caused by the tangential force component, $F_{\theta}$, is readily derived. Conversely, calculating the energy dissipation attributable to the radial force component, $F_r$, necessitates integrating to average the dissipation energy across the integral term.
\begin{align}
    \int_{0}^{2 \pi} \frac{\sin^2 \theta (1+cos\theta)^6}{\dot\psi- \frac{(1+cos\theta)^2}{(1-e^2)^{3/2}} n}   d\theta
\end{align}
As the expression $\frac{(1+\cos\theta)^2}{(1-e^2)^{3/2}}$ nears $\dot\psi$, the limit tends towards infinity, resembling the behavior of the Dirac delta function's derivative at $\pi/2$ subtracted by its derivative at $3\pi/2$. With varying eccentricity, the divergence point of the Dirac delta function shifts slightly. Despite this, using approximate values at different $\dot\psi$ and $n$ levels does not fulfill our goal of formulating a tidal evolution equation that encompasses all orders of eccentricity.

The occurrence of overshoot is attributed to the equation's lack of completeness in terms of eccentricity order inclusion, as seen in Section 2, where even low eccentricity maintains angular momentum conservation. Therefore, while a fully encompassing expression for all orders of eccentricity remains elusive, we can enhance the tidal evolution equation by extending the expansion. By starting with \cite{Zahn1966Les} and expanding to the fourth order, we mitigate the "overshoot" issue, especially when the initial eccentricity is set to $0.3$, thereby refining our understanding of tidal potentials.


\begin{equation}
\begin{aligned}
U=\frac{G M_{c}}{a}\left(\frac{r}{a}\right)^{2} \cdot & {\left[-\frac{1}{2} P_{2}(\cos \theta) \cdot\left(1+3 e \cos \omega t+e^{2}\left(\frac{3}{2}+\frac{9}{2} \cos 2 \omega t\right)\right.\right.} \\
&\left.+e^{3}\left(\frac{27}{8} \cos \omega t+\frac{53}{8} \cos 3 \omega t\right)+e^{4}\left(\frac{15}{8}+\frac{7}{2} \cos 2 \omega t+\frac{77}{8} \cos 4 \omega t\right)\right) \\
&+\frac{1}{4} P_{2}^{2}(\cos \theta) \cdot\left(\left(\cos (\omega t-2 \phi)+e\left(\frac{7}{2} \cos (3 \omega t-2 \phi)-\frac{1}{2} \cos (\omega t-2 \phi)\right)\right.\right.\\
&+e^{2}\left(\frac{17}{2} \cos (4 \omega t-2 \phi)-\frac{5}{2} \cos (2 \omega t-2 \phi)\right) \\
&+e^{3}\left(\frac{1}{16} \cos (\omega t-2 \phi)+\frac{1}{48} \cos (\omega t+2 \phi)\right.\\
&\left.+\frac{845}{48} \cos (5 \omega t-2 \phi)-\frac{123}{16} \cos (3 \omega t-2 \phi)\right) \\
&+e^{4}\left(\frac{533}{16} \cos (6 \omega t-2 \phi)+\frac{1}{24} \cos (2 \omega t+2 \phi)\right.\\
&\left.\left.\left.-\frac{115}{16} \cos (4 \omega t-2 \phi)+\frac{13}{16} \cos (2 \omega t-2 \phi)\right)\right)\right]+O\left(e^{5}\right)
\end{aligned}
\end{equation}
In this analysis, the spherical coordinates $r$, $\theta$, and $\phi$ are utilized, consistent with the conventions outlined in \cite{zahn1977tidal}. Our goal is to determine the perturbative acceleration, adopting their approach by setting the secondary body as the origin point, with polar coordinates represented as $(r, O, f - \psi t)$ \cite{zahn1977tidal}. Drawing from solar system dynamics, the mean anomaly is defined as follows:
\begin{align}
\begin{aligned}
      &f=\omega t+2e\sin \omega t+\frac{5}{4}e^2+e^3(\frac{13}{12}\sin 3\omega t-\frac{1}{4} \sin \omega t)  \\ 
    &\phantom{ f=M+2e\sin M+\frac{5}{4}e^2}+e^4 (\frac{103}{96}\sin 4\omega t-\frac{11}{24} \sin 2\omega t)+O(e^5)
\end{aligned}
\end{align}
Following the methodology \cite{zahn1977tidal}, we express the perturbing acceleration as a Fourier series in terms of $\omega t$, categorizing the odd terms under $R$ and the even terms under $S$. Consequently, we articulate the perturbing acceleration up to the fourth order, with the radial component denoted as $R$ and the tangential component as $S$.
\begin{align}
\begin{aligned}
     &R=-\frac{9}{4}\frac{G m_s}{R^2}(\frac{R_p}{a})^7 (\frac{a}{r})^4  \\
        &\phantom{R} \cdot[e \sin \omega t (\epsilon_2^{10}+\frac{1}{4}\epsilon_2^{12}-4\epsilon_2^{22}+\frac{7}{2}\epsilon_2^{32}) \\
        &\phantom{R} + e^2 \sin 2\omega t (\frac{9}{4}\epsilon_2^{20}+\epsilon_2^{12}-\frac{5}{2}\epsilon_2^{22}-7\epsilon_2^{32}+\frac{17}{2}\epsilon_2^{42}]
            \end{aligned}\\
\begin{aligned}
        &S=-\frac{3}{2}\frac{G m_s}{R^2}(\frac{R_p}{a})^7(\frac{a}{r})^4 \\  
        &\phantom{S} \cdot [\epsilon_2^{22}+e\cos \omega t (-\frac{1}{2}\epsilon_2^{12}+\frac{7}{2}\epsilon_2^{32})   \\
        &\phantom{S} +e^2(\epsilon_2^{12}-\frac{13}{2}\epsilon_2^{22}+7\epsilon_2^{22})   \\
        &\phantom{S}+e^2\cos 2\omega t (-\epsilon_2^{12}+4\epsilon_2^{22}-7\epsilon_2^{32}+\frac{17}{2}\epsilon_2^{42})  \\
        &\phantom{S}+e^3\cos\omega t (\frac{27}{16}\epsilon_2^{12}-5\epsilon_2^{22}-\frac{165}{16}\epsilon_2^{32}+17\epsilon_2^{42})  \\
        &\phantom{S}+e^4 (-\epsilon_2^{12}+\frac{57}{4}\epsilon_2^{22}-39\epsilon_2^{32}+\frac{221}{8}\epsilon_2^{42})]
\end{aligned}
    \end{align}
In the term $S$, components of orders $3$ and $4$ are present, yet upon averaging, their effective order exceeds $4$. For instance, averaging a term like $e^3 \cos 3\omega t$ results in contributions of eccentricity order $6$.

Adhering to the principles outlined in \cite[Chapter 2]{murray2000solar}, we examine the variations in the semi-major axis, eccentricity, and spin velocity, which are influenced by the torque $-M_c r S$:
\begin{align}
\begin{aligned}
 &\frac{d}{d t}a=\frac{2}{n\sqrt{1-e^2}}[e\sin{f}R+(1-e^2)\frac{a}{r}S] \\
        &\frac{d}{d t}e=\frac{\sqrt{1-e^2}}{n a}[\sin{f}R+(\cos{f}+\cos{E})S] \\
        &\frac{d}{d t}\dot \psi=-\frac{M_2}{C} r S 
\end{aligned}
\end{align}
Now, we substitute the components $R$ and $S$.
    \begin{align}
    \begin{aligned}
           &\frac{d a}{d t}=-\frac{3}{n}\frac{M_p+M_s}{M_p}\frac{G M_s}{R_p^2}(\frac{R_p}{a})^7  \\
        &\phantom{\frac{d a}{d t}}\cdot [\epsilon_2^{22}+e^2(\frac{3}{4}\epsilon_2^{10}+\frac{1}{8}\epsilon_2^{12}-5\epsilon_2^{22}+\frac{147}{8}\epsilon_2^{32}) \\
        &\phantom{\frac{d a}{d t}}+e^4(\frac{32}{27}\epsilon_2^{10}+\frac{81}{16}\epsilon_2^{20}+\frac{115}{64}\epsilon_2^{12}-6\epsilon_2^{22}-\frac{3579}{64}\epsilon_2^{32}-\frac{527}{4}\epsilon_2^{42})+O(e^6)]\\
        &\frac{d e}{d t}=-\frac{3}{4}\frac{e}{n}\frac{M_p+M_s}{M_p}\frac{G M_s}{R_p^3}(\frac{R_p}{a})^8  \\
        &\phantom{\frac{d e}{d t}}
        \cdot[\frac{3}{2}\epsilon_2^{10}-\frac{1}{4}\epsilon_2^{12}-\epsilon_2^{22}+\frac{49}{4}\epsilon_2^{32} \\
        &\phantom{\frac{d a}{d t}}+e^4(\frac{3}{16}\epsilon_2^{10}+\frac{81}{8}\epsilon_2^{20}+\frac{119}{32}\epsilon_2^{12}-\frac{45}{2}\epsilon_2^{22}-\frac{919}{32}\epsilon_2^{32}+119\epsilon_2^{42})+O(e^6)]\\
        &\frac{d}{d t}(\dot \psi_p)=\frac{3}{2}\frac{G M_s^2}{C_p R_p}(\frac{R_p}{a})^6  \\
        &\phantom{\frac{d}{d t}(I\Omega)} \cdot [\epsilon_2^{22}+e^2(\frac{1}{4}\epsilon_2^{12}-5\epsilon_2^{22}+\frac{49}{4}\epsilon_2^{32}) \\
        &\phantom{\frac{d a}{d t}}+e^4(-\frac{1}{16}\epsilon_2^{12}+\frac{63}{8}\epsilon_2^{22}-\frac{861}{16}\epsilon_2^{32}+\frac{289}{4}\epsilon_2^{42})+O(e^6)]
    \end{aligned}
    \end{align}
We verify the accuracy of this equation within the $\Delta t$ model at the fourth order by substituting $\epsilon_2^{lm} = k_2 (ln - m\dot{\psi}) \Delta t$.
\begin{align}
\begin{aligned}
       &\frac{d\dot\psi_i}{d t}=-\frac{3G}{C_i }k_{2i}\triangle t_i {M_j}^2 (\frac{R_i^5}{a^6})[(1+\frac{15}{2}e^2+\frac{105}{4}e^4)\psi_i-(1+\frac{27}{2}e^2+\frac{573}{8}e^4)n+O(e^6)]  \\
    &\frac{d X}{d t}=\frac{6G}{\mu X^7}k_{2p}\triangle t_p {M_s}^2(\frac{1}{R_p} )^3[(1+\frac{27}{2}e^2+\frac{573}{8}e^4)(\frac{\psi_p}{n}+A_{\triangle t}\frac{\psi_c}{n})   \\
    &\phantom{\frac{d X}{d t}=\frac{6G}{\mu X^7}k_{2p}\triangle t_p {M_s}^2(\frac{1}{R_p})^3+{}}-(1+23e^2+180e^4)(1+A_{\triangle t})+O(e^4)] \\
    &\frac{d e}{d t}=\frac{27e G}{\mu X^8}K_{2p}\triangle t_p {M_s}^2 (\frac{1}{R_p})^3[\frac{11}{18}(1+\frac{13}{2}e^2)(\frac{\dot{\psi_p}}{n}+A_\triangle t \frac{\dot{\psi_s}}{n}) \\
    &\phantom{\frac{d e}{d t}=\frac{27e G}{\mu X^8}K_{2p}\triangle t_p {M_s}^2 (\frac{1}{R_p})^3}-(1+\frac{41}{4}e^2)(1+A_\triangle t)+O(e^2)]
\end{aligned}
\end{align}
Having examined the tidal equation within the $\Delta t$ model, we now aim to extend this analysis to the Q model and proceed with its simulation.
    \begin{align}
    \label{Q_exactly}
    \begin{aligned}
         &\frac{d \dot{\psi_i}}{d t}=-\frac{3 G M_j^2}{2 C_i}\frac{k_{2 i}}{Q_i}\frac{R_i^5}{a^6}[sgn(\dot\psi-n)+e^2 D_i+e^4 G_i+O(e^6)]  \\
        &\frac{1}{a}\frac{d a}{d t}=3 n \frac{k_{2p}}{Q_p}\frac{M_c}{M_p}(\frac{R_p}{a})^5 
        [sgn(\dot{\psi}-n)+A_Q sgn(\dot\psi_c-n)   \\
        &\phantom{\frac{1}{a}<\frac{d a}{d t}>=3 n \frac{k_{2p}}{Q_p}\frac{M_c}{M_p}(\frac{R_p}{a})^5}+e^2(E_p+A_Q E_c)+e^4(H_p+A_Q H_c)+O(e^6)]  \\
        &\frac{1}{e}\frac{d e}{d t}=n \frac{k_{2p}}{Q_p} \frac{M_c}{M_p}(\frac{R_p}{a})^5[F_p+A_Q F_c + e^2 (J_p+A_Q J_c) + O(e^4)]  
    \end{aligned}
    \end{align}
\begin{table}[!htbp]
    \centering
    \caption{coefficient of \eqref{Q_exactly}}
    \begin{tabular}{c c c c c c c}
    \hline
     $\dot\psi/n$ & $D_i$ & $E_i$ & $F_i$ & $G_i$ & $H_i$ & $J_i$\\ \hline
     $>2$ & $15/2$ & $51/4$ & $57/8$ & $105/4$ & $2103/32$ & $2937/64$  \\ 
     $=2$ & $15/2$ & $51/4$ & $57/8$ & $-46$ & $-2113/32$ & $-2775/64$  \\ 
     $>3/2 and < 2$ & $15/2$ & $51/4$ & $57/8$ & $-473/4$ & $-6329/32$ & $-8487/64$  \\ 
     $=3/2$ & $-19/4$ & $-45/8$ & $-33/16$ & $-1031/16$  & $-9079/64$ & $-14217/128$  \\
     $>1 and < 3/2$ & $-17$ & $-24$ & $-45/4$  & $-85/8$ & $-1375/16$ & $-2865/32$  \\
     $=1$ & $-12$ & $-19$ & $-21/2$ & $-37/2$ & $-1279/16$ & $-2325/32$  \\
     $>1/2 and <1$ & $-7$ & $-14$ & $-39/4$  & $-211/8$ & $-1183/16$ & $-1785/32$  \\
    \end{tabular}
        
    \label{tab:my_label}
\end{table}
In the presented figure, we conducted simulations based on evolution \eqref{eqn: 789}, which approximates the second order. Notably, this figure reveals the presence of an unnecessary overshoot in the simulation results.
\begin{figure}[!htbp]
    \centering
    \includegraphics[width=15cm]{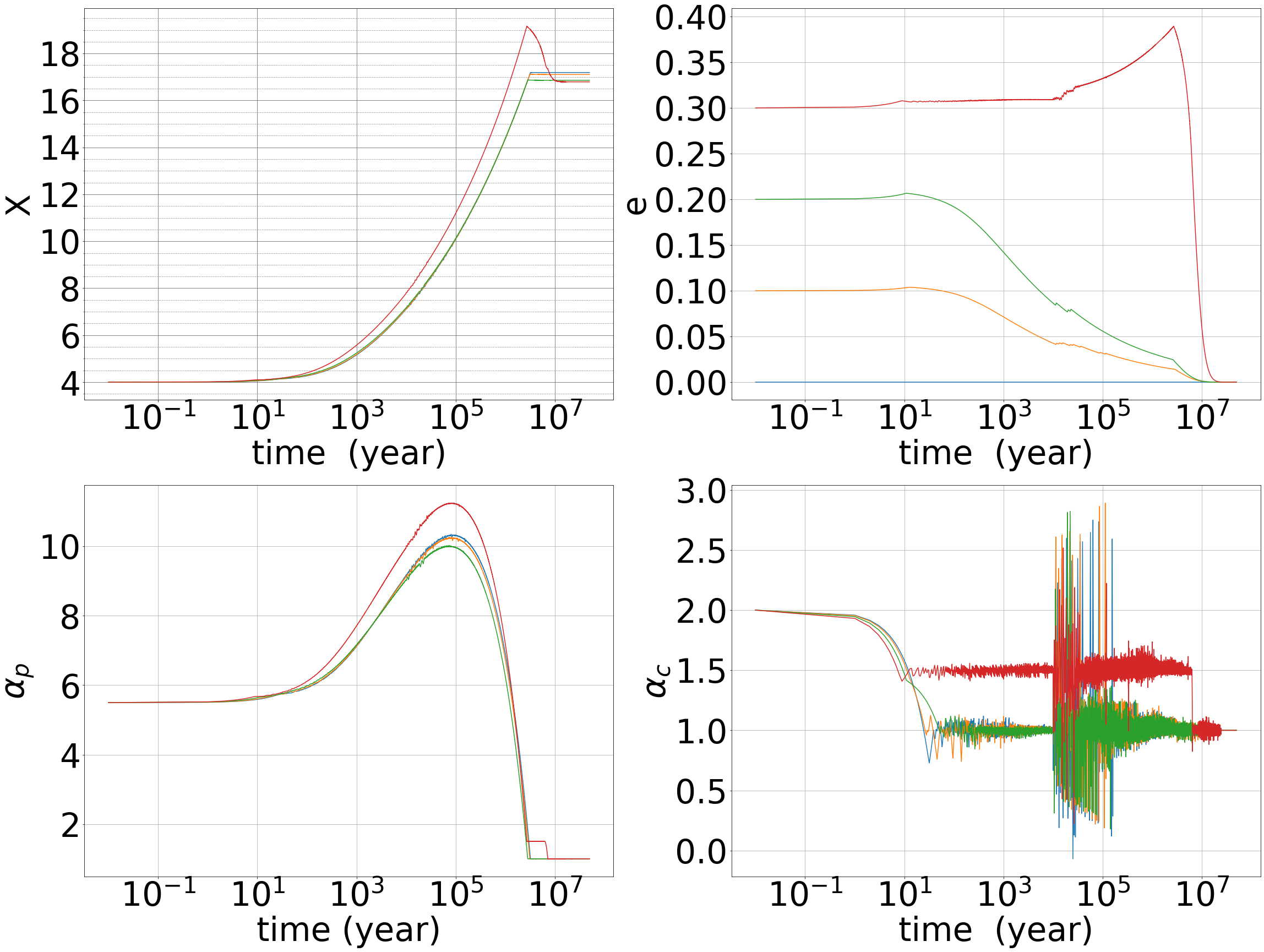}
    \caption{Tidal evolution at Q model follows \eqref{eqn: 123} with $\epsilon_2^{lm}=\frac{k_2}{Q} sgn(l n -m \psi)$, and its initial state is $\alpha_p=5.5$, $\alpha_c=2$, $X=4$, and $e=0.1, 0.2$, and $0.3$. The $A_Q$ is $1.15$.}
    \label{Q_2}
\end{figure}
Subsequently, we align the figure with \eqref{Q_exactly}, where I have computed the subsequent order term of eccentricity. This allows us to simulate eccentricities of $0$, $0.1$, $0.2$, and $0.3$, effectively resolving the issue of the unnecessary overshoot.
\begin{figure}[!htbp]
    \centering
    \includegraphics[width=15cm]{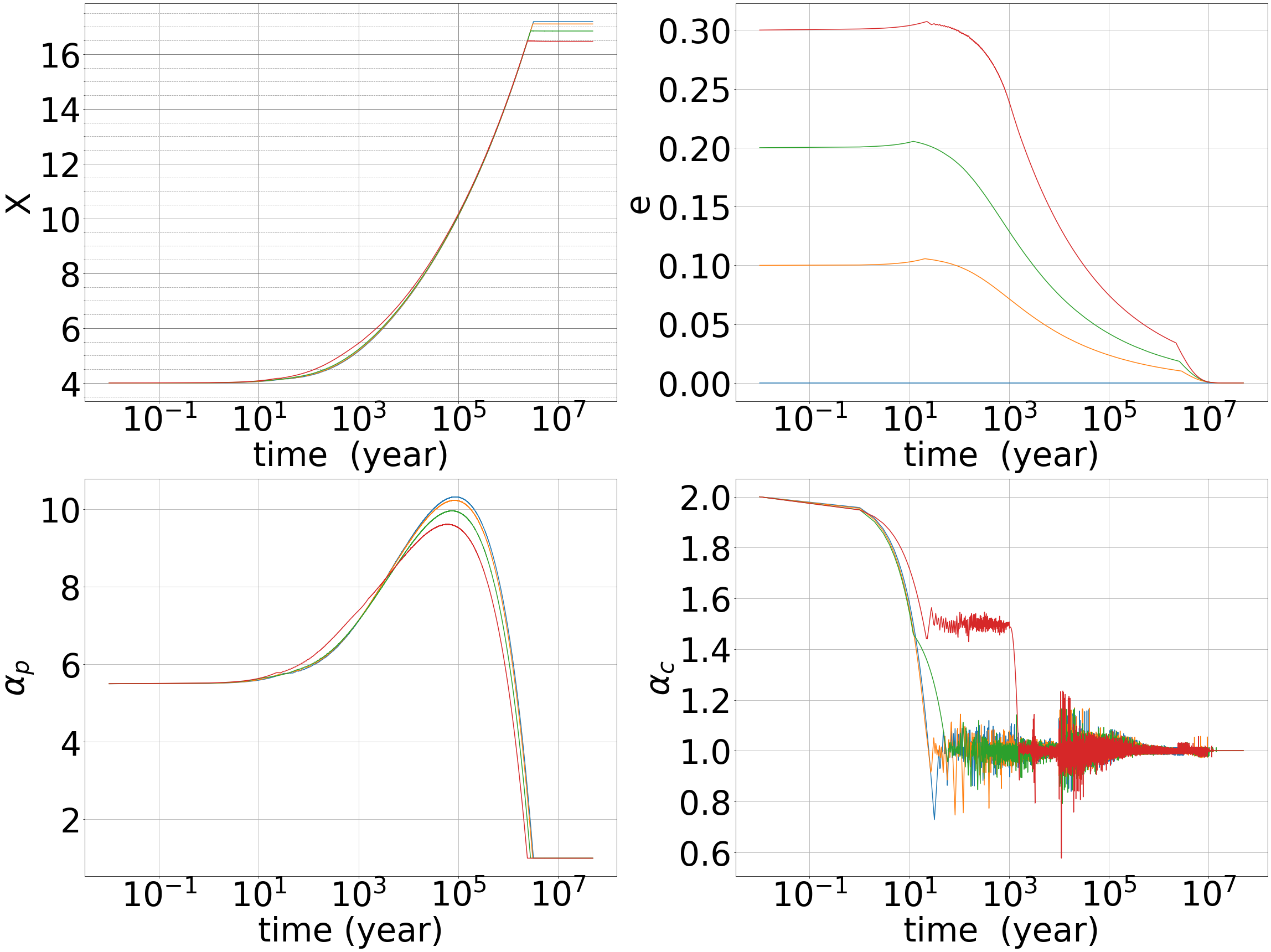}
    \caption{This figure illustrates the tidal evolution within the Q model as described by evolution \eqref{Q_exactly}, starting from an initial state with $\alpha_p=5.5$, $\alpha_c=2$, a semi-major axis to Pluto's radius ratio $X$ of $4$, and eccentricities $e$ of $0.1$, $0.2$, and $0.3$. The parameter $A_Q$ is set to $1.15$.}
    \label{Q_all}
\end{figure}
However, this approach does not effectively address the issue of higher eccentricities within the Q model. As the eccentricity increases, the current expression fails to adequately manage these higher values. As discussed in Section 2, when utilizing tidal evolution to narrow down the range of possible initial states, eccentricities in the range of $0.2$ to $0.5$ are of particular interest. While initial values of $\alpha_p$ also influence eccentricity, our focus remains on significant eccentricity values for studying other celestial bodies with non-zero eccentricities. Therefore, the Q model may not be the most suitable for such simulations, prompting a preference for the $\Delta t$ model in these instances.
\begin{figure}[H]
    \centering
    \includegraphics[width=15cm]{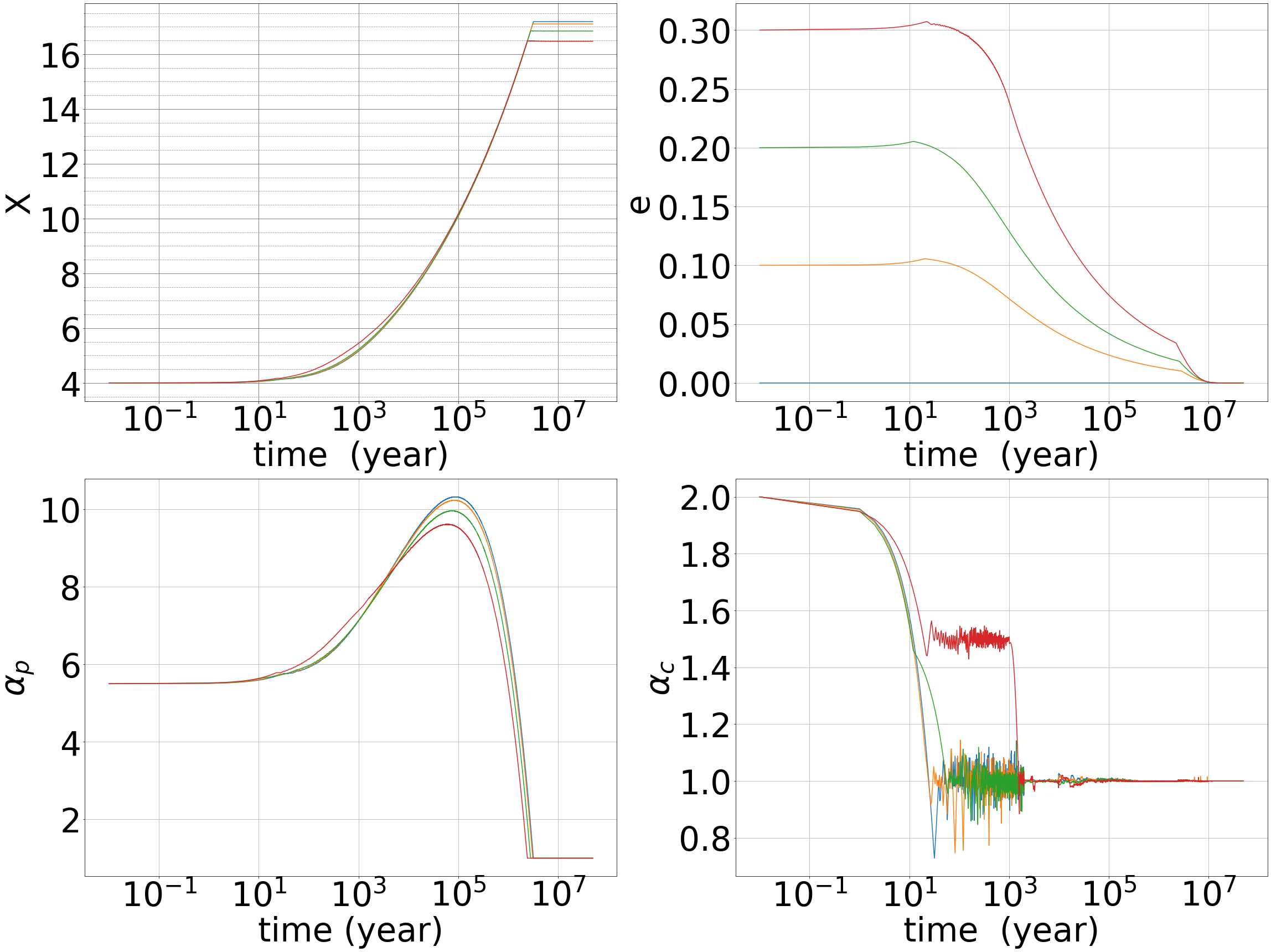}
    \caption{This figure represents the averaged outcomes of 300 data points from Fig.\ref{Q_all} observed after 2000 years.}
    \label{Q_continue}
\end{figure}
In Fig. \ref{Q_continue}, the simulation of the Q model exhibits discontinuities. To approximate the sign function, we selected data points where the absolute value is less than $10^{-4}$. In the later stages of the simulation, $\alpha_c$ exhibits rapid fluctuations. The subsequent figure presents the result of averaging $300$ data points, taken after $2000$ years.

\section{Tidal Evolution Including Inclination}
Despite Charon's current inclination nearing zero, the initial inclination remains a significant factor, akin to the role of eccentricity in tidal evolution, where the endpoint typically results in zero inclination. It is thus imperative to examine the implications of initial inclination values on tidal evolution.

We revert to the $\Delta t$ model for this analysis. Although there are multiple methods to explore tidal evolution, the $\Delta t$ model proves to be more effective in our simulations, offering a direct representation rather than an approximate expression of tidal evolution. We adopt the equations from \cite{mignard1980evolution} to incorporate inclination into the tidal evolution framework, as follows:
\begin{align}
\begin{aligned}
     &\frac{1}{n}\frac{d\dot\psi_i}{d t}=-\frac{3G {M_j}^2}{C_i R_i}k_{2i}\triangle t_i  (\frac{R_i}{a})^6  [f_1(e)\frac{\psi_i}{n}cos I-f_2(e^2)] \\
    &\frac{1}{a}\frac{d a}{d t}=\frac{6G}{\mu R_p^3}k_{2p}\triangle t_p {M_c}^2(\frac{R_p}{a} )^8[f_2(e^2)(\frac{\psi_p}{n}cos I+A_{\triangle t}\frac{\psi_c}{n})   \\
    &\phantom{\frac{d X}{d t}=\frac{6G}{\mu X^7}k_{2p}\triangle t_p {M_s}^2(\frac{1}{R_p})^3+{}}-f_3(e^2)(1+A_{\triangle t})] \\
    &\frac{1}{e}\frac{d e}{d t}=\frac{27G}{\mu R_p^3}K_{2p}\triangle t_p {M_s}^2 (\frac{R_p}{a})^8[f_4(e^2)\frac{11}{18}(\frac{\dot{\psi_p}}{n}cos I+A_\triangle t \frac{\dot{\psi_s}}{n}) \\
    &\phantom{\frac{d e}{d t}=\frac{27e G}{\mu X^8}K_{2p}\triangle t_p {M_s}^2 (\frac{1}{R_p})^3}-f_5(e^2)(1+A_\triangle t)]\\
    &\frac{d I}{d t}=\sum_i-\frac{3G}{C_i\ R_i}k_{2i}\triangle t_i {M_s}^2(\frac{R_i}{a})^6\frac{n}{\dot\psi_p} sin I [f_2(e^2)  \\
    &\phantom{\frac{d I}{d t}=-\frac{3G {M_s}^2}{C_p\ R_p}}
    -\frac{1}{2}(cos I-\frac{C_i}{\mu}(\frac{1}{a})^2(1-e^2)^{-1/2}\frac{\dot\psi_i}{n})\frac{\dot\psi_i}{n} f_1(e^2)]
\end{aligned}
\end{align}
where
\begin{align*}
    &f_1(e^2)=(1+3e^2+\frac{3}{8}e^4)/(1-e^2)^{9/2} \\
    &f_2(e^2)=(1+\frac{15}{2}e^2+\frac{45}{8}e^4+\frac{5}{16}e^6)/(1-e^2)^{6} \\
    &f_3(e^2)=(1+\frac{31}{2}e^2+\frac{255}{8}e^4+\frac{185}{16}e^6+\frac{25}{64}e^8)/(1-e^2)^{15/2} \\
    &f_4(e^2)=(1+\frac{3}{2}e^2+\frac{1}{8}e^4)/(1-e^2)^{5} \\
    &f_5(e^2)=(1+\frac{15}{4}e^2+\frac{15}{8}e^4+\frac{5}{64}e^6)/(1-e^2)^{13/2} 
\end{align*}

\begin{figure}[H]
    \centering
    \includegraphics[width=15cm]{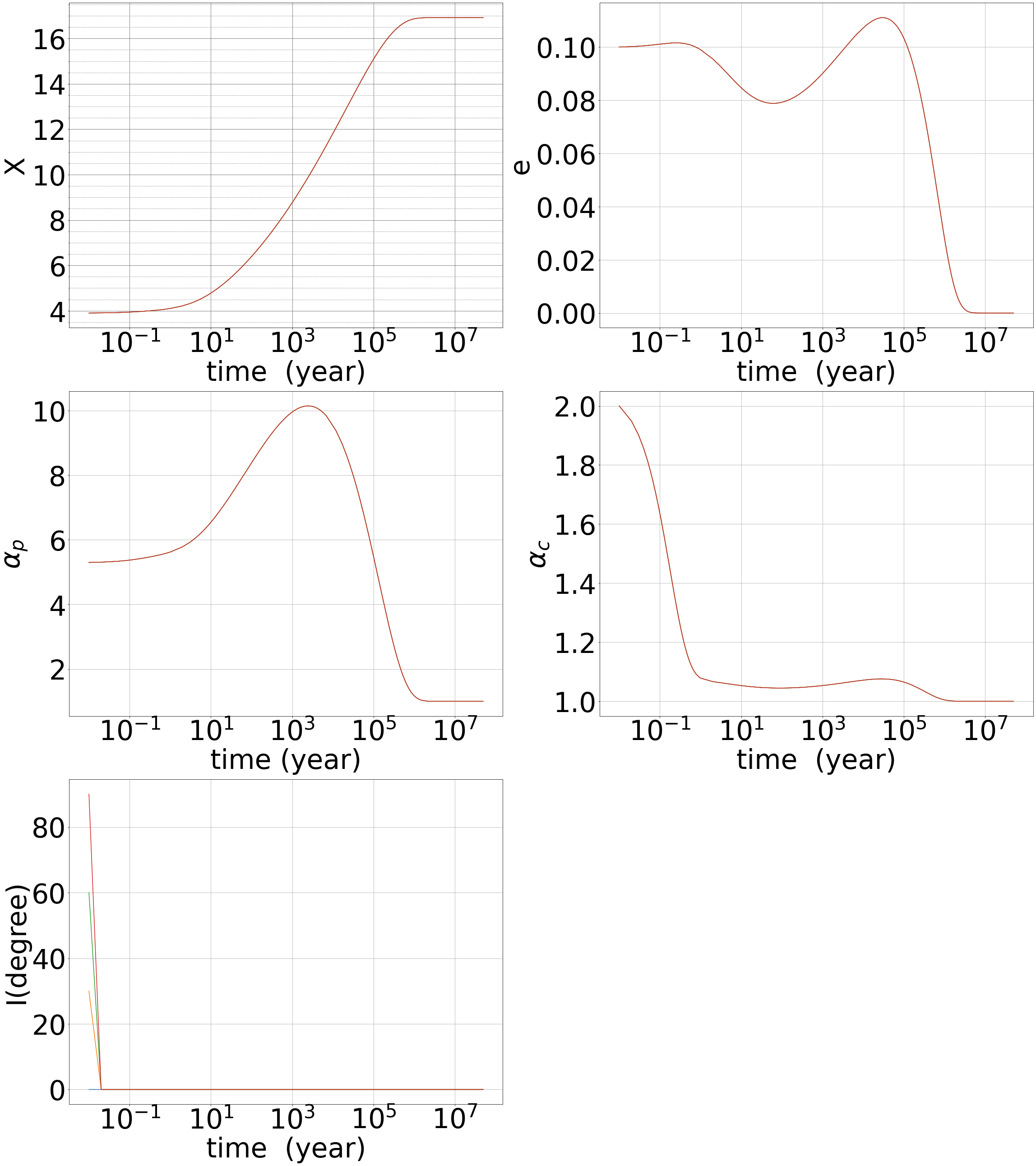}
    \caption{This figure demonstrates that varying initial inclinations have no significant impact on tidal evolution. The initial conditions are set with $\alpha_p=5.3$, $\alpha_c=2$, a semi-major axis to Pluto's radius ratio $X$ of $3.9$, and eccentricity $e$ of $0.1$ while testing inclinations at $0^{\circ}$, $30^{\circ}$, $60^{\circ}$, and $90^{\circ}$.}
    \label{fig: 21}
\end{figure}

\begin{figure}[H]
    \centering
    \includegraphics[width=15cm]{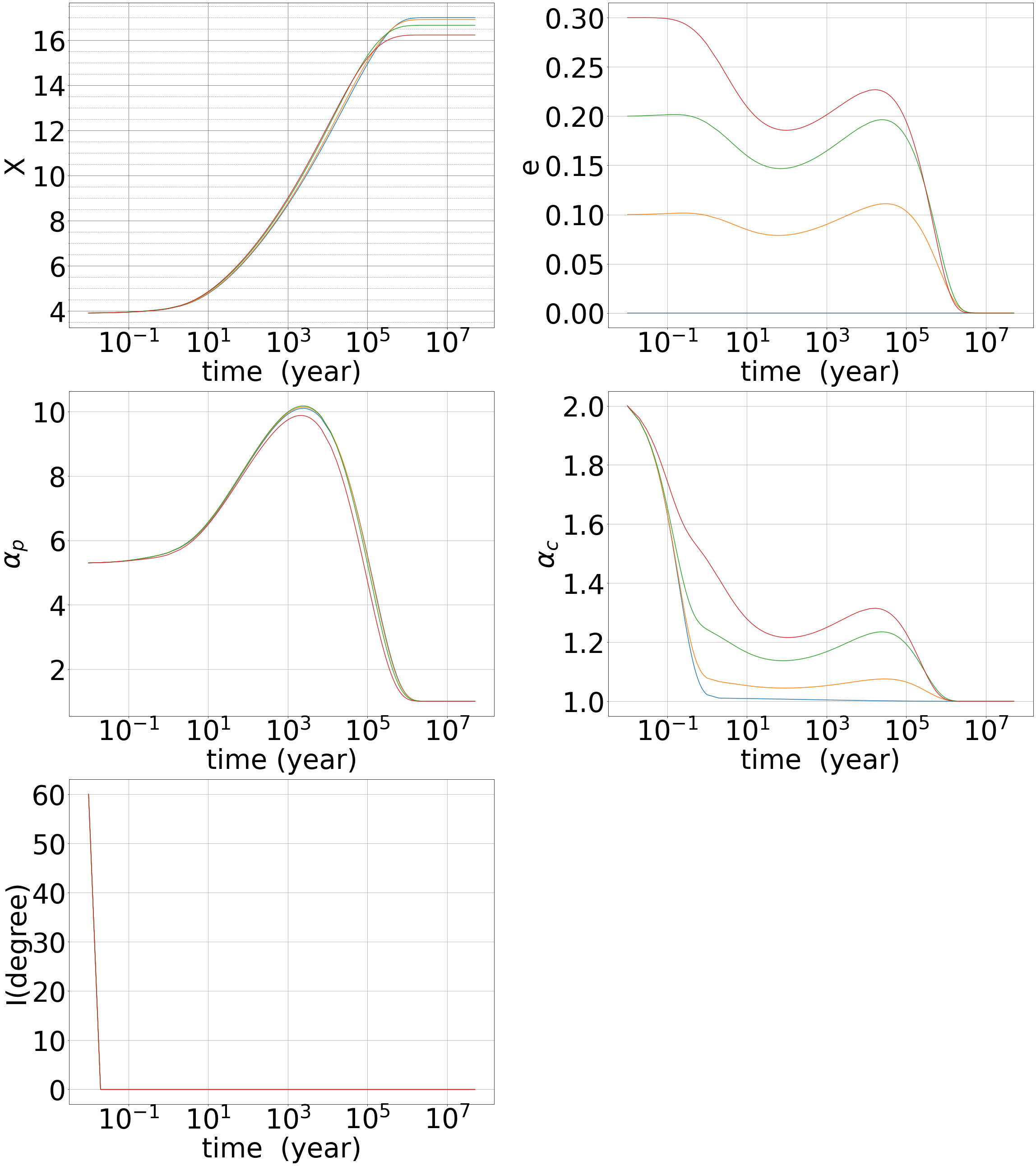}
    \caption{This figure illustrates that variations in initial eccentricity do not influence the tidal evolution process. The initial conditions include an inclination of $60^{\circ}$, $\alpha_p=5.3$, $\alpha_c=2$, a semi-major axis to Pluto's radius ratio $X$ of $3.9$, with eccentricities set at $0$, $0.1$, $0.2$, and $0.3$.}
    \label{fig: 22}
\end{figure}

Fig.\ref{fig: 21} illustrates that inclination diminishes rapidly and has a negligible impact on tidal evolution, a trend that persists even when the inclination is adjusted to $90^{\circ}$. Furthermore, Fig.\ref{fig: 22} corroborates that varying eccentricities do not alter this outcome, affirming that inclination does not significantly influence tidal evolution.

\section{Discussion}
\subsection{Effects of $A_{\triangle t}$ and $A_Q$ Parameters}
The parameters $Q_p$ and $\triangle t_p$ significantly impact the pace of tidal evolution, not only shifting the peak of $\alpha_p$ but also adjusting the overall trajectory of tidal evolution. Conversely, the parameters $A_{\triangle t}$ and $A_Q$ introduce perturbations into this evolutionary process. Higher values of $A_{\triangle t}$ and $A_Q$ result in a rapid decline in both eccentricity and $\alpha_c$, leading to a somewhat predictable outcome. In contrast, lower values of $A_{\triangle t}$ and $A_Q$ slow down the decay rates of eccentricity and $\alpha_c$, introducing notable perturbations and complexity into the system's evolution. This observation aligns with \cite{cheng2014complete}, which indicates that exceedingly low values of $A_{\triangle t}$ and $A_Q$ prevent the eccentricity from settling to zero. Furthermore, these perturbations play a role in the occurrence of overshoots; as eccentricity increases, so does the necessity for a larger orbital semi-major axis, making overshoots an inevitable aspect of the system's evolution.

\begin{figure}[!htbp]
    \centering
    \includegraphics[width=15cm]{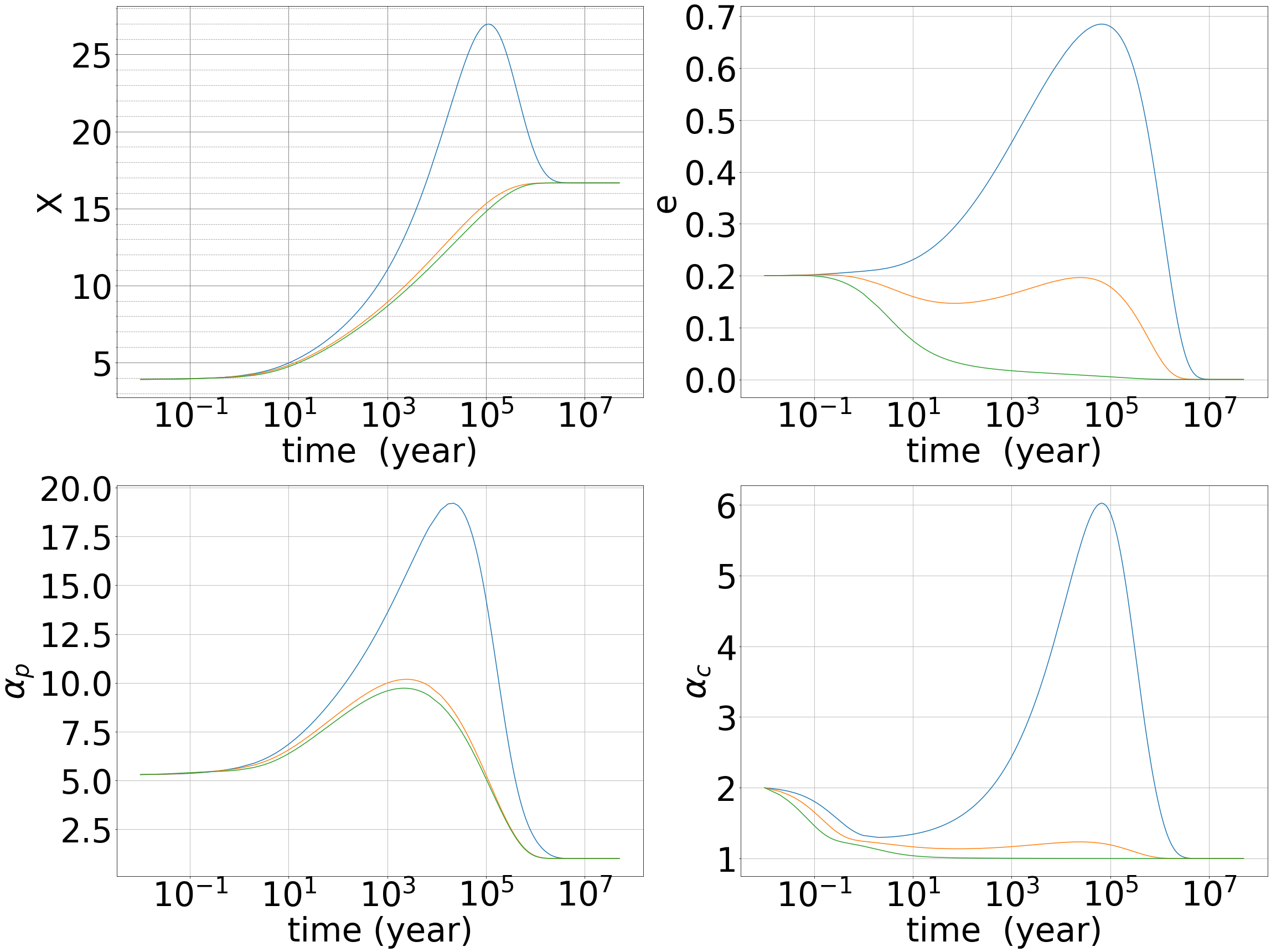}
    \caption{This figure highlights the significant role of $A_{\Delta t}$ in tidal evolution. The red, green, and blue lines represent $A_{\Delta t}$ values of $20$, $10$, and $5$, respectively. The initial conditions are set with $\alpha_p=5.3$, $\alpha_c=2$, a semi-major axis to Pluto's radius ratio $X$ of $3.9$, and eccentricity $e=0.2$.}
    \label{fig: 23}
\end{figure}

This leads to a fascinating observation: tidal forces can accelerate the spin of celestial bodies and result in a larger orbital semi-major axis, coinciding with a period of high eccentricity. This phenomenon is depicted in Fig. \ref{fig: 23}.

\begin{figure}[!htbp]
    \centering
    \includegraphics[width=15cm]{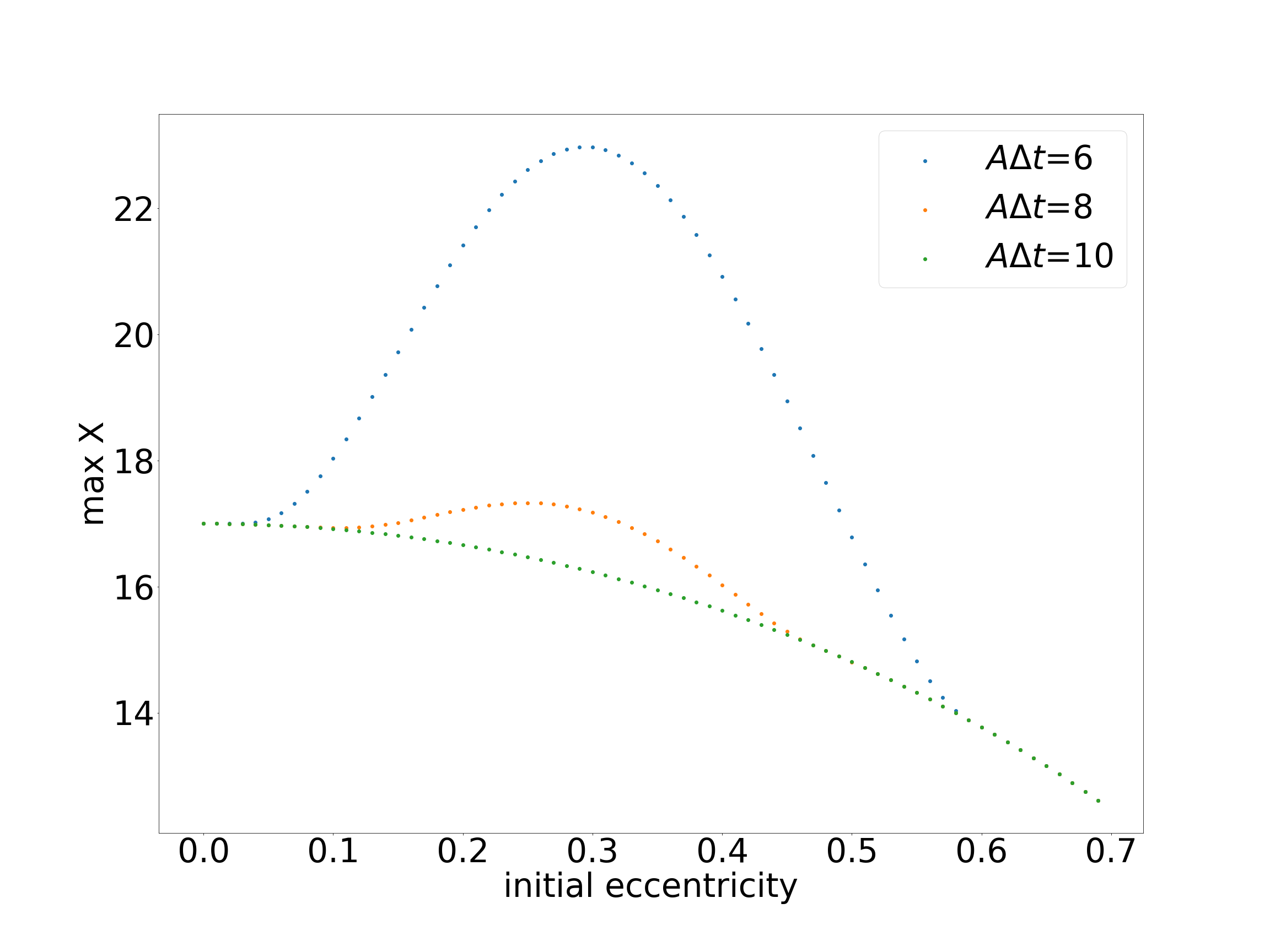}
    \caption{This figure illustrates the maximum semi-major axis $X$ for various initial eccentricities using the $\Delta t$ model. The x-axis represents different initial eccentricities, while the initial conditions include $\alpha_p=5.3$, $\alpha_c=2$, and a semi-major axis to Pluto's radius ratio $X$ of $3.9$.}
    \label{fig: 24}
\end{figure}

Building on the final point from Section 2, having a larger eccentricity doesn't always mean it has a bigger effect on how things change over time. In Fig. \ref{fig: 24}, we explore how the maximum distance $X$ relates to the starting conditions for different settings of $A_{\Delta t}$. When $A_{\Delta t}$ is high, we don't see this effect, but it's still something worth looking into. This kind of situation is tricky to study whether we use the Q model or the $\Delta t$ model, and it's not something we usually see in planets and their moons. In most cases with planets and moons, we don't consider the moon's effect to be significant. This makes it an interesting point to start studying double star systems, where a high $A_{\Delta t}$ can cause this effect to appear.

In Section 2, we confirmed the findings of \cite{cheng2014complete}, showing that various eccentricities uphold the conservation of angular momentum. We highlighted initial conditions at specific eccentricities and Charon's spin velocities that closely adhere to this conservation principle. We also examined the evolution of the orbital semi-major axis and its relationship with $\alpha_p$, which is significantly influenced by $A_{\Delta t}$ and $A_Q$. However, identifying suitable $A_{\Delta t}$ and $A_Q$ values for the Pluto-Charon system proved challenging, indicating that these parameters critically affect tidal evolution. This analysis underscores the relevance of tidal evolution models for studying binary star systems affected by tidal forces.

Section 3 addressed issues from Section 2 related to the truncation of eccentricity orders, employing a more precise Q model equation to mitigate overshoot tendencies. Yet, lower $A_Q$ values led to convergence issues, suggesting the $\Delta t$ model as a more reliable descriptor for tidal evolution due to its accuracy at lower $A_Q$ values.

We also explored the impact of inclination, finding that it diminishes rapidly compared to other parameters and appears to have a minimal effect on tidal evolution.

Finally, we investigated special conditions related to overshoot, examining various eccentricities to determine the maximum values of X and $\alpha_p$ at low $A_{\Delta t}$. This revealed scenarios with high semi-major axes, spin velocities, and eccentricities, indicating significantly altered stellar characteristics, including temperature variations due to the wide range of orbital semi-major axes. This highlights the importance of considering tidal evolution in binary star systems with low $A_{\Delta t}$ values.

Future research could focus on identifying binary star systems with low $A_{\Delta t}$ values, where tidal forces play a crucial role before other parameters are considered.

\bibliographystyle{IEEEtran}
\bibliography{references}

\end{document}